\documentclass[english,aps,prl,superscriptaddress,twocolumn,showpacs,footinbib]{revtex4-1}
\usepackage{graphicx}  
\usepackage{dcolumn}   
\usepackage{bm}        
\usepackage{amssymb}   
\usepackage{babel}
\usepackage{verbatim}
\usepackage{amsmath}
\usepackage{setspace}
 \usepackage{epstopdf}
\graphicspath{{figures/}}
\usepackage{siunitx}
\usepackage{mathtools}
\DeclarePairedDelimiter{\ceil}{\lceil}{\rceil}

\begin{document}

\widetext

\title{Identifying and decoupling many-body interactions in spin ensembles in diamond}

\author{D. Farfurnik}
\affiliation{Racah Institute of Physics, The Hebrew University of Jerusalem, Jerusalem 9190401, Israel}
\affiliation{The Center for Nanoscience and Nanotechnology, The Hebrew University of Jerusalem, Jerusalem 9190401, Israel}

\author{Y. Horowicz}
\affiliation{Racah Institute of Physics, The Hebrew University of Jerusalem, Jerusalem 9190401, Israel}

\author{N. Bar-Gill}
\affiliation{Racah Institute of Physics, The Hebrew University of Jerusalem, Jerusalem 9190401, Israel}
\affiliation{The Center for Nanoscience and Nanotechnology, The Hebrew University of Jerusalem, Jerusalem 9190401, Israel}
\affiliation{Dept. of Applied Physics, Rachel and Selim School of Engineering, The Hebrew University of Jerusalem, Jerusalem 9190401, Israel}


\date{\today}

\begin{abstract}
In this work, we simulate the dynamics of varying density quasi-2D spin-ensembles in solid state systems, focusing on the Nitrogen-Vacancy (NV) centers in diamond. We consider the effects of various control sequences on the averaged dynamics of large ensembles of spins, under a realistic ``spin-bath" environment. We reveal that spin-locking is efficient for decoupling spins initialized along the driving axis, both from coherent dipolar interactions, and from the external spin-bath environment, when the driving is two orders of magnitude stronger than the relevant coupling energies. Since the application of standard pulsed dynamical decoupling (DD) sequences leads to strong decoupling from the environment, while other specialized pulse sequences can decouple coherent dipolar interactions, such sequences can be used to identify the dominant interaction type. Moreover, a proper combination of pulsed decoupling sequences could lead to the suppression of both interaction types, allowing additional spin manipulations. Finally, we consider the effect of finite-width pulses on these control protocols, and identify improved decoupling efficiency with increased pulse duration, resulting from the interplay of dephasing and coherent dynamics.
    
\end{abstract}

\maketitle

\section{I. introduction}
The studies of many-body dynamics of spin ensembles in the solid state have attracted significant attention. In particular, ensembles of negatively-charged nitrogen-vacancy (NV) centers were recently used for the demonstration of many-body depolarization dynamics \cite{Choi2017,Kucsko2017}. A proper application of microwave (MW) control sequences on the spin ensemble of interest may lead to a variety of applications in quantum sensing and quantum information processing. For example, a proper modification of traditional sequences in NMR such as WAHUHA \cite{Waugh1968,Rhim1973} and MREV \cite{Mansfield1971}, could result in engineered Hamiltonians for the interacting spins within the ensemble. Such engineered Hamiltonians could pave the way toward the creation of non-classical states, e.g. spin squeezed-states,  which could eventually lead to magnetic sensing beyond the shot-noise limit \cite{Rey2008, Cappellaro2009}. Studies of such control sequences on interacting spin ensembles, in the presence of noise and for various parameter regimes, are still lacking.
\paragraph{}
In this work, we use a cluster-based simulation method to estimate the dynamics of a quasi-2D ensemble consisting of more than 400 spins under various control sequences. The simulations consider a realistic environment, consisting of a ``spin-bath" noise, representing the typical scenario of dense ensembles of NV centers in diamond. Our analysis identifies techniques for controllably decoupling specific interactions, and clarifies the effects of finite pulse durations originating from the interplay of different interaction sources.   
\paragraph{}
The electronic structure of the negatively-charged NV center has a spin-triplet ground state, in which the $m_s=\pm 1$ sublevels experience a zero-field splitting ($\sim 2.87$ GHz) from the $m_s = 0$ sublevel due to spin-spin interactions. Application of an external static magnetic field along the NV symmetry axis Zeeman shifts the $m_s=\pm 1$ levels. If MW driving is applied at a frequency $\omega_0$ resonant with the $m_s= 0\leftrightarrow+1$ transition (for example), this spin manifold can be treated as a two-level subspace of the spin-triplet \cite{Taylor2008}. 
\section{II. Interactions and decoupling}
The effective Hamiltonian representing dipolar interactions within such an ensemble is \cite{Kucsko2017}:
\begin{equation}
H_{dipolar}=\sum_{ij} w_{ij}[\vec{\sigma}_i \cdot \vec{\sigma}_j-2\sigma^z_i\sigma^z_j],
\label{eq:dipolar}
\end{equation}
with $\omega_{i,j}=\frac{J_0}{r_{ij}^3}$, where $\sigma^{x,y,z}_{i,j}$ are the Pauli spin operators, $r_{ij}^3$ is the distance between spins $i$ and $j$, and $J_0\approx 52$ MHz$\cdot$nm$^3$. The environment of NV centers corresponds to a ``spin-bath", typically dominated by $^{13}$C nuclear and nitrogen paramagnetic spin impurities, which create time-varying random local magnetic fields in the crystal \cite{Pham2012,Acosta2009,deSousa2009,BarGill2012}. In typical samples, the interaction between such a bath and the spins of interest is modeled as an Ornstein-Uhlenbeck (OU) process $B(t)$ \cite{deSousa2009,BarGill2012}, with a typical correlation time $\tau_c$ of the bath, and coupling strength between the bath and the spins of interest $b$. The effect of the resulting fluctuating fields on the spin ensemble (and assuming that these fields are uniform within the measurement volume) can be modeled as the interaction Hamiltonian
\begin{equation}
H_{bath}=B(t)\sum_{i}\sigma^z_{i}. 
\label{eq:bath}
\end{equation}
The amplitude of the random bath noise as a function of time can be simulated by an exact algorithm \cite{Gillespie1996}:
\begin{equation}
B(t+\Delta t)=B(t)e^{-\Delta t/\tau_c}+\frac{b}{2} n \sqrt{1-e^{-2\Delta t/\tau_c}},
\label{eq:OU}
\end{equation}
where $n$ is a randomly generated number from a normal distribution with mean 0 and standard deviation of 1.

\paragraph{}
For many decades, dynamical decoupling (DD) MW sequences were used in NMR for decoupling spin interactions, and thus controlling their dynamics \cite{Hahn1950,Meiboom1958,Mansfield1971,Rhim1973,Gullion1990,Khodjasteh2005,Ryan2010,Hirose2012}. In the case of spin ensembles, the simplest decoupling method method from the bath is to apply a continuous driving at the resonant frequency $\omega_0$ \cite{Hirose2012} which (in the rotating frame) takes the form    
\begin{equation}
H_{SL}=\Omega \sum_{i}\sigma^x_{i},
\label{eq:spinlock}
\end{equation}
where all the spins are assumed to be driven by the same strength $\Omega$. If all spins are initialized along the driving (``x") axis, the driving overcomes the effects of frequency terms in the spin-bath with frequencies lower than $\Omega$ (``spin-locking"). This enhances the fidelity of the initial state with time up to a timescale usually referred to as $T_{1\rho}$ which, for sufficiently high $\Omega$, is typically limited by phononic interactions and experimental imperfections \cite{Cai2012,Hirose2012,Farfurnik2017cont}, but may also be limited by dipolar interactions in extremely-dense ensembles \cite{Choi2017}. Another method for decoupling from the bath is the repetitive application of resonant $\pi$-pulses. In the simplest implementation, the Carr-Purcell-Meiboom-Gill (CPMG) sequence, all pulses are applied along the initialization (``x") axis \cite{Meiboom1958}, while other sequences for overcoming pulse imperfections are available \cite{Khodjasteh2005,Ryan2010,Wang2012a,Farfurnik2015}. Another useful decoupling sequence is WAHUHA \cite{Waugh1968,Rhim1973,Mansfield1971}, consisting of four unequally spaced resonant $\pi/2$ pulses, which was designed to decouple collective dipolar interactions between spin-1/2 particles at times determined by the sequence length. Since the NV dipolar Hamiltonian \eqref{eq:dipolar} differs from the spin-1/2 dipolar Hamiltonian by the term $-\sigma^z_i\sigma^z_j$, the WAHUHA sequence is expected to decouple the dipolar interactions only partially. In the average Hamiltonian picture, the remaining isotropic Hamiltonian $\sum \omega_{ij}\vec{\sigma}_i \cdot \vec{\sigma}_j$ conserves the total angular momentum $J^2$, providing advantages toward the creation of high-fidelity non-classical states over the scenario of spin-1/2 dipolar coupling in which the effective Hamiltonian is just 0 \cite{Rey2008, Cappellaro2009} (Appendix A), highlighting the importance of the current studies of spin dynamics under these unique interactions. In this work, we simulate the effect of continuous and pulsed decoupling techniques on the dynamics of a spin ensemble under spin-bath and collective dipolar interactions with different strengths. We demonstrate procedures for distinguishing between these types of interactions, for decoupling from them, and identify the physical mechanisms underlying effects resulting from finite pulse durations.  

\section{III. Simulations}
We use a cluster-based simulation \cite{Maze2008b} (Appendix B) to estimate the dynamics of a quasi-2D spin ensemble consisting of more than 400 spins, interacting by dipolar interactions according to eq. \eqref{eq:dipolar}. All spins in the ensemble are initialized along the $x$ axis, and the spin polarization along the initialization axis (expectation value $\langle S_x \rangle$) is extracted as a function of time (Fig. \ref{fig:dipolar}). Since simulating the evolution of the total density matrix in such a large ensemble is impractical, we follow the general idea of dividing the ensemble into independent clusters \cite{Maze2008b}, and perform the simulation by combining the results for small clusters in the following way (Appendix B): First, we obtain a ``representative probability distribution" for describing dipolar interaction strengths in a typical experimental scenario, by randomly generating the positions of 464 spins inside a circle with a surface of $\approx 4.5$ $\SI{}{\micro\meter}^2$. Afterwards, we use this distribution for evaluating the exact dynamics of small spin clusters: a few (4-10) spins are generated under the distribution, and their dynamics under the Hamiltonian \eqref{eq:dipolar} is simulated exactly.  We repeat this spin generation process with newly generated clusters, to take into account all possible scenarios for the proximity between spins within clusters, until the number of realizations is sufficient for the convergence of the resulting dynamics. In such a way, each realization samples the dynamics within a certain cluster consisting of a few spins, and the averaged dynamics of all realizations provides insights into the combined dynamics of the whole ensemble. Similar procedures could be done for 3D-ensembles, with the only difference being that the distribution of spin couplings should take into account their different orientations.
\subsection{A. Dipolar Dynamics}
\begin{figure}[!t]	
	\includegraphics[width=1\columnwidth]{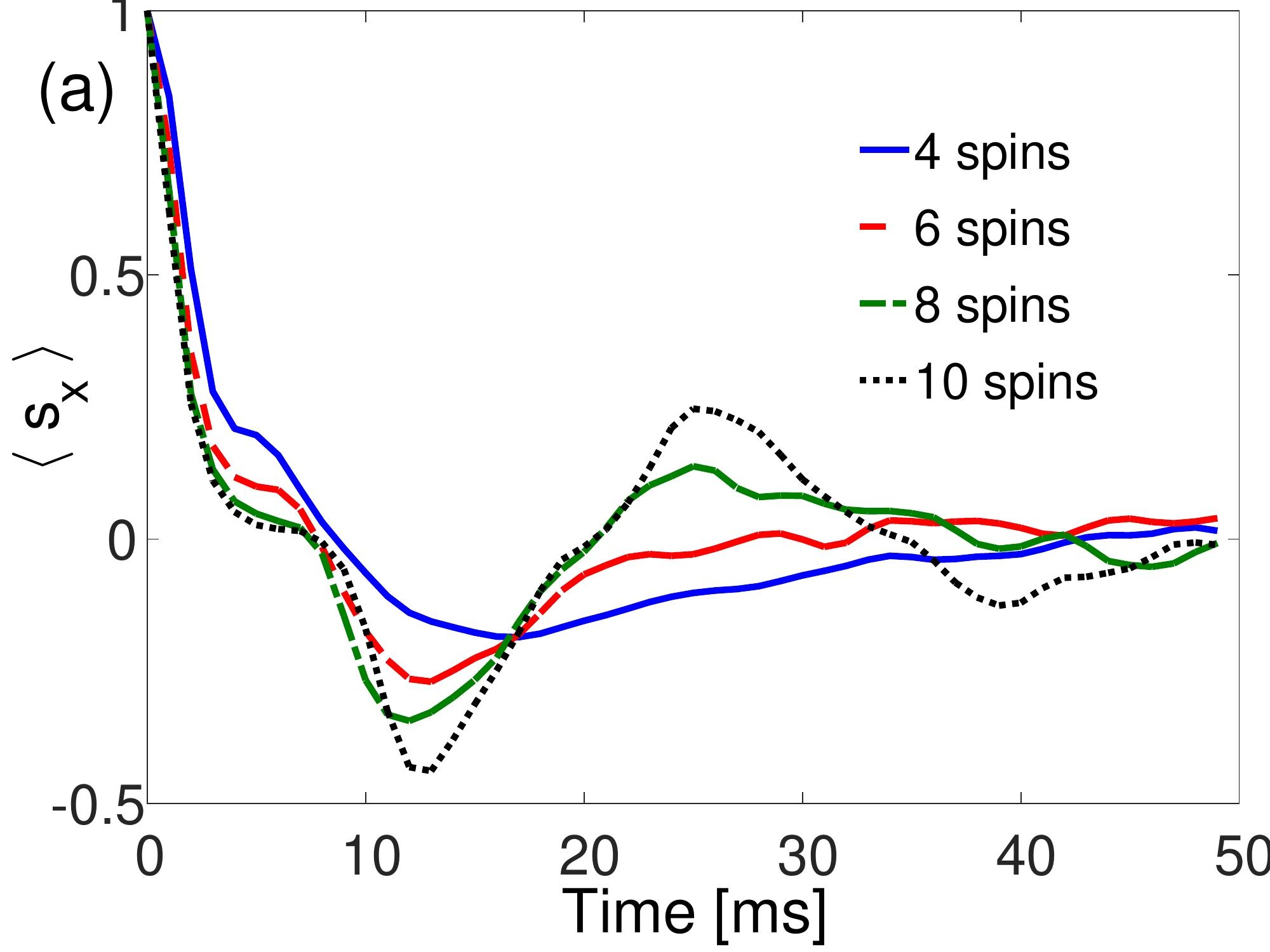} 
	\includegraphics[width=1\columnwidth]{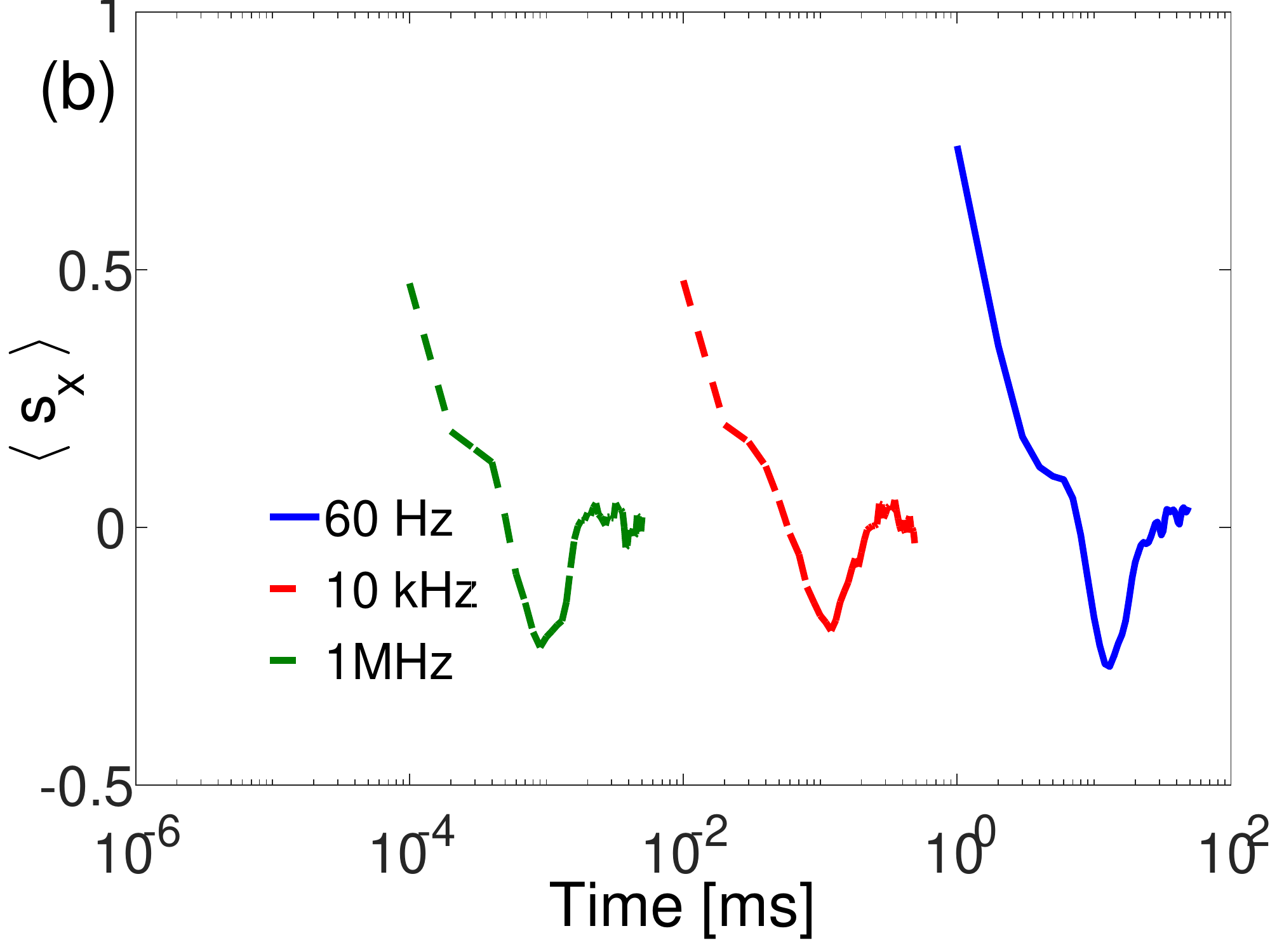} 
	\caption{(Color online) Cluster-based simulations of the dynamics of a spin ensemble, dominated by internal dipolar interactions. (a) Spin concentration of $10^{10}$ $\SI{}{\centi\meter}^{-2}$ within a $\approx 4.5$ $\SI{}{\micro\meter}^2$ measurement surface, representing 464 spins with a typical interaction strength of $\sim 60$ Hz. Different curves represent different numbers of interacting spins taken into account in a cluster. (b) Varying typical dipolar strengths, using clusters of six spins: 60 Hz and 10 kHz typical interactions  within a $\approx 4.5$ $\SI{}{\micro\meter}^2$ measurement surface (464 and 9980 spins), and 1 MHz within a $\approx 0.46$ $\SI{}{\micro\meter}^2$ measurement surface (9980 spins).}
	\label{fig:dipolar}
\end{figure}

The Spin dynamics simulations of NV ensembles under the Hamiltonian \eqref{eq:dipolar} are shown in Fig. \ref{eq:dipolar}.
Although our simulation method does not take into account collective phenomena such as extended, long-range many-body correlations, our detailed convergence analysis (Appendix B) indicates that such phenomena do not significantly affect the spin dynamics of driven systems studied in this work. The significant frequencies (4,12 and 20 times the typical interaction strength) contributing to the spin dynamics (Fig. \ref{fig:dipolar}) can be predicted from an analytical expression for all-to-all equal interactions, with $25\%$ accuracy of evaluating the typical interaction strength. Additional convergence analysis emphasizes that the choice of six-spin clusters provides a converged quantitative estimate for spin dynamics even for large numbers of spins (Appendix B). As a result, and in order to optimize run times, the simulations in this work utilize clusters with six spins only. Qualitatively, the resulting decay structures are similar for different dipolar interaction strengths, with only the decay timescales changing by the same factor as the interaction strength ratios (stronger interactions lead to a faster decay). Small quantitative variations remain, though,  due to the difference in the spin generation probability distributions for different spin concentrations (Appendix C) [Fig. \ref{fig:dipolar}(b)]. 
\subsection{B. Spin-Locking}
\begin{figure}[]	
	\includegraphics[width=0.85\columnwidth]{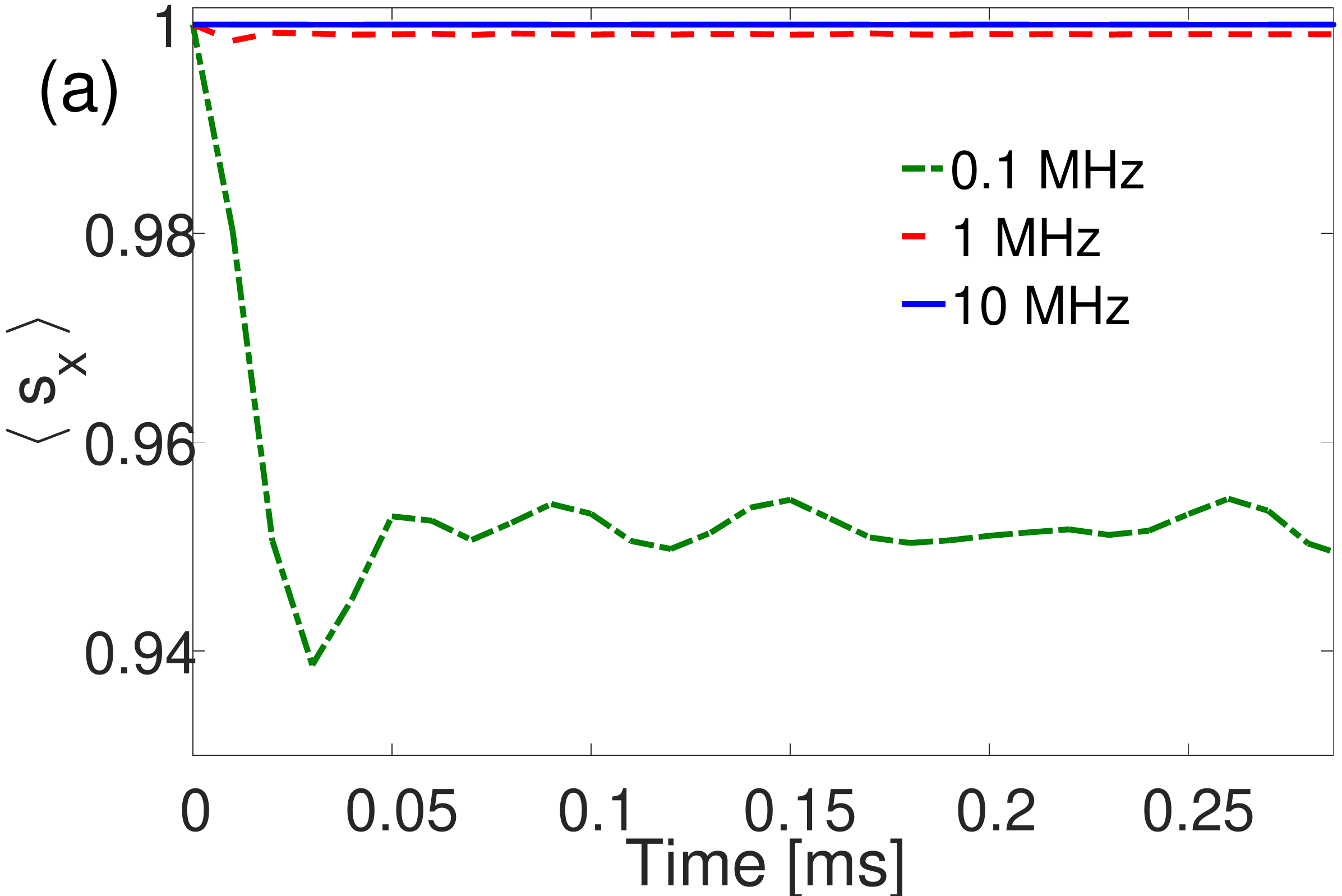} 
	\includegraphics[width=0.87\columnwidth]{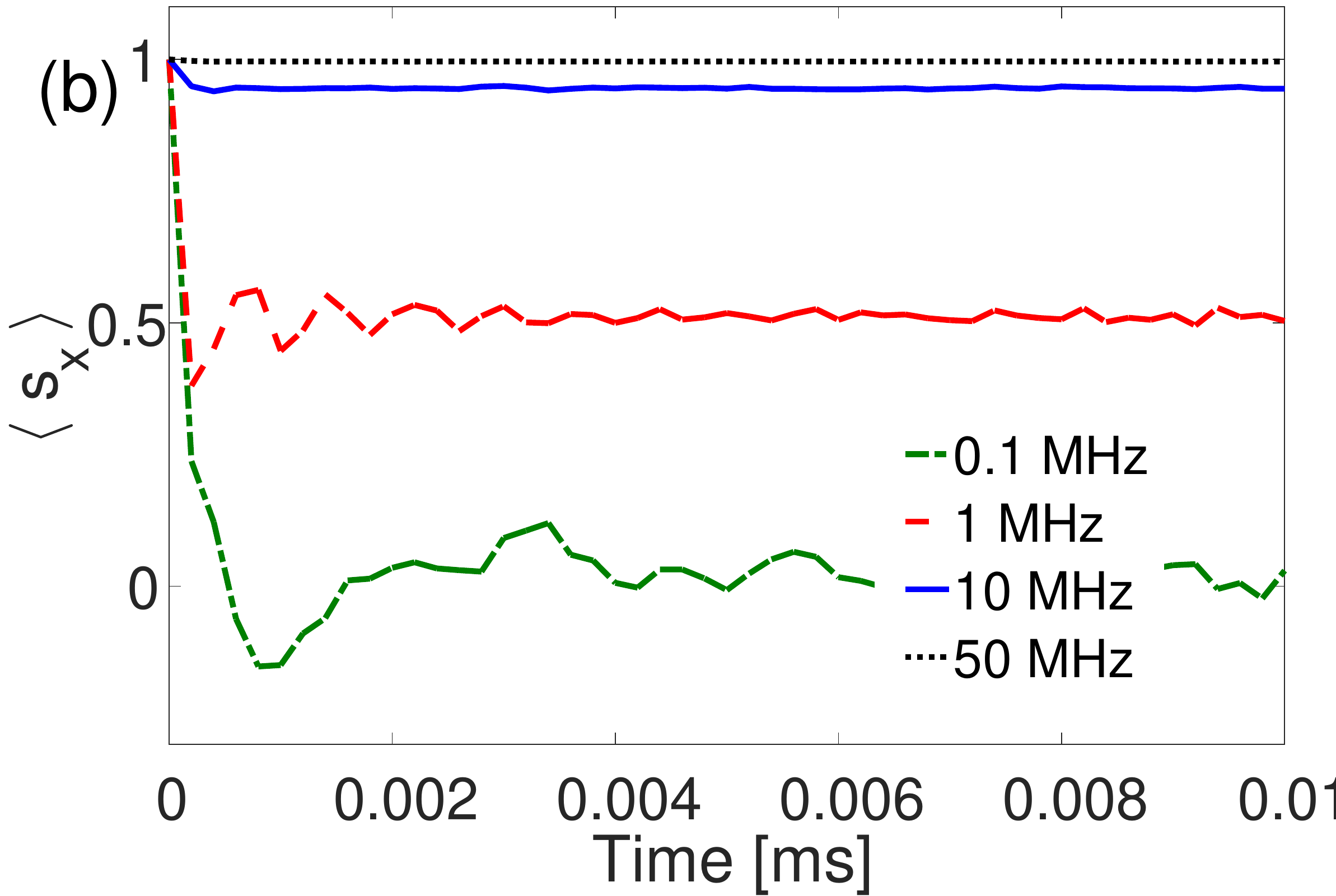} 
	\includegraphics[width=1\columnwidth]{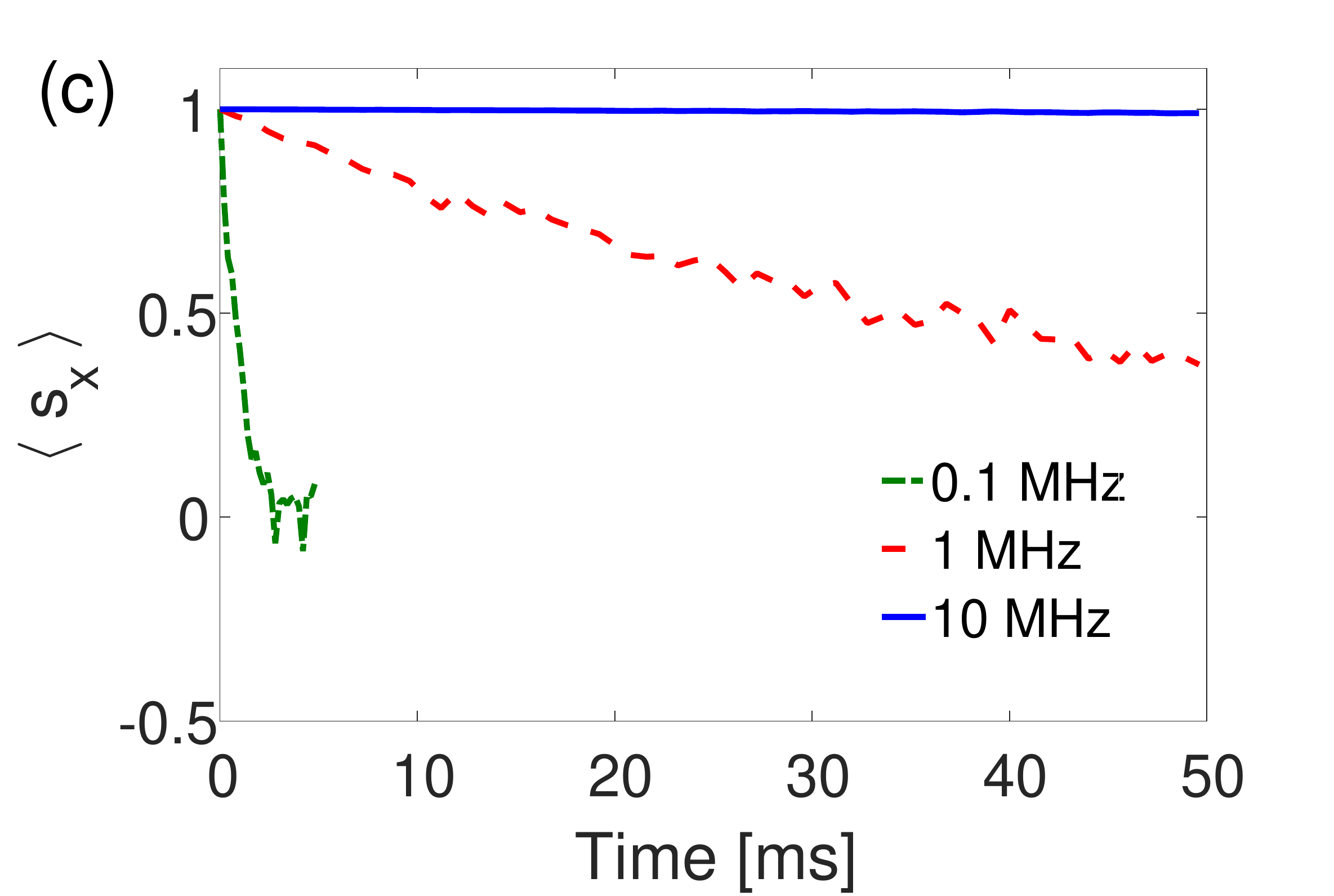} 
	\caption{(Color online) Cluster-based simulations of the spin dynamics of an ensemble consisting of 9980 spins, under spin-lock driving at different intensities, for dipolar interactions of (a) 10 kHz (a $\approx 4.5$ $\SI{}{\micro\meter}^2$ measurement surface) and  (b) 1 MHz (a $\approx 0.46$ $\SI{}{\micro\meter}^2$ measurement surface) within the ensemble. For a dipolar interaction of 60 Hz, the evolution was unity even for the weakest examined spin-lock intensity. (c) Effect of spin-locking on an ensemble with dipolar interactions of 60 Hz (464 spins), in a spin-bath environment with $\tau_c=5$ $\mu$s correlation time and $b=20$ kHz coupling strength to the ensemble, averaging only 5 dipolar realizations.}
	\label{fig:spinlock}
\end{figure}

Next, we consider the effect of spin-lock driving applied along the initialization axis, on the dynamics of the ensemble. Our simulations considering different spin concentrations, which correspond to different dipolar coupling strengths (Fig. \ref{fig:spinlock}), demonstrate that without including the effect of the bath [Fig. \ref{fig:spinlock} (a),(b)], when the driving intensity is two orders of magnitude stronger than the dipolar coupling, the dipolar interactions are fully decoupled. In particular, for a dipolar coupling of $\sim 60$ Hz, which could be achieved for NV ensembles using standard CVD procedures and TEM irradiation \cite{Farfurnik2015,Farfurnik2017}, even a driving as weak as 0.1 MHz results in a complete preservation of the initial state (Appendix C). 
These results agree with our theoretical model: consistent with the results of Fig. \ref{fig:spinlock}, an analytical expression for the spin dynamics within a six-spin cluster (Appendix B) correctly predicts full decoupling under spin-lock two orders of magnitude stronger than the typical dipolar strength. Moreover, for spin-locking one order of magnitude stronger than the typical interaction strength, the simulations in Fig. \ref{fig:spinlock} result in the converged baseline of 0.945, in agreement with the conserved population 0.9375 predicted from this theoretical model (Appendix B).   
\paragraph{}
In a more realistic scenario, one has to take into account the additional effects of the spin-bath environment interacting with the spin ensemble. By considering the OU model within a time $T$, the simulation of spin dynamics under such an environment with $\tau_c=5$ $\SI{}{\micro\second}$ and $b=20$ kHz using an exact algorithm according to eq. \eqref{eq:bath} leads to a free induction decay (FID) time of $T_2^*\approx 0.5$ ms, consistent with theoretical calculations [$T_2^*=1/(b^2\tau_c)=0.5$ ms] \cite{deSousa2009}. Since the simulation of OU processes requires evolution in small time increments $\Delta t$ and averages over noise realizations, it is significantly more time-consuming, and therefore for the combined simulation under spin-bath and dipolar interactions only five realizations of the dipolar interactions were considered. Although the exact dynamics varies with the specific interactions in the generated cluster (Appendix C), the total trend remains clear: Fig. \ref{fig:spinlock}(c) demonstrates that by considering the evolution of a spin ensemble with a dipolar strength of 60 Hz, spin-lock driving two orders of magnitude stronger than the interactions with the bath ($5$ $\SI{}{\micro\second}$ $\leftrightarrow 0.2$ MHz) decouples both the spin-bath and dipolar interactions within the ensemble, resulting in a unity evolution on a timescale of $50$ ms. Such long coherence times can be experimentally observed at cold temperatures ($\sim 77$ K), for which the longitudinal relaxation time ($T_1$) is much longer \cite{Jarmola2012,Farfurnik2015}. 
\subsection{C. Pulsed Decoupling}
\begin{figure}[!t]	
	\includegraphics[width=0.85\columnwidth]{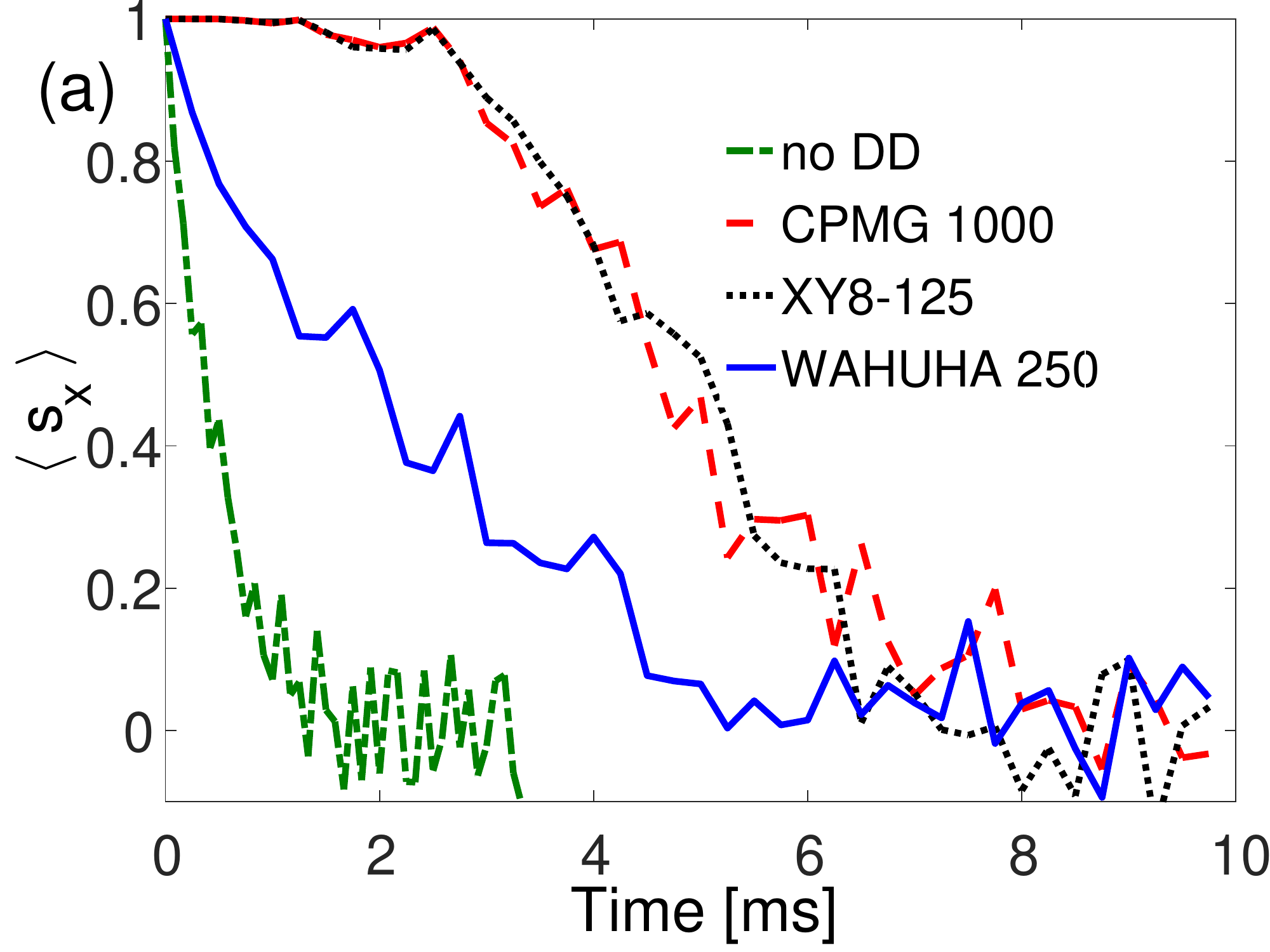} 
	\includegraphics[width=0.87\columnwidth]{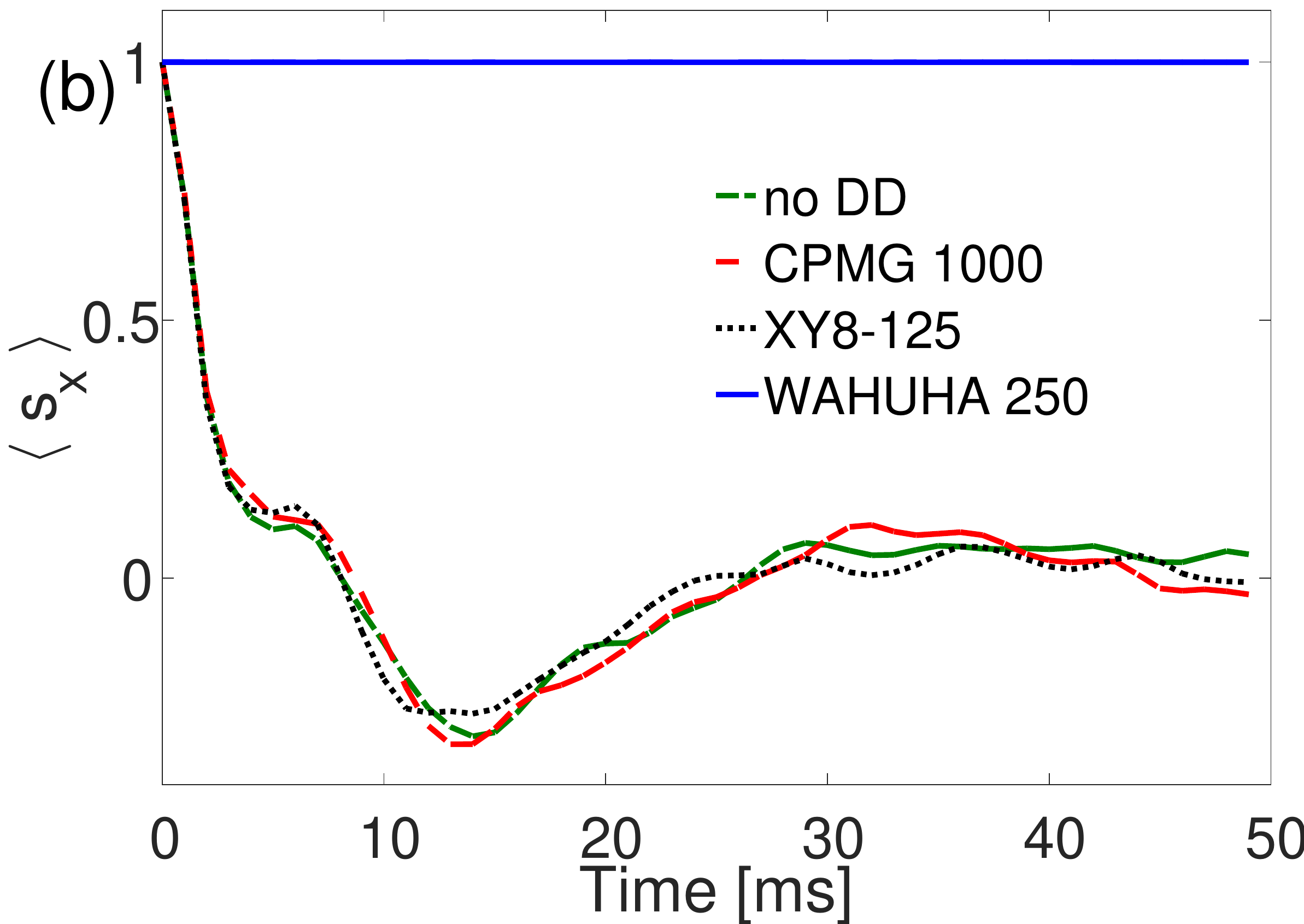}
	\includegraphics[width=1\columnwidth]{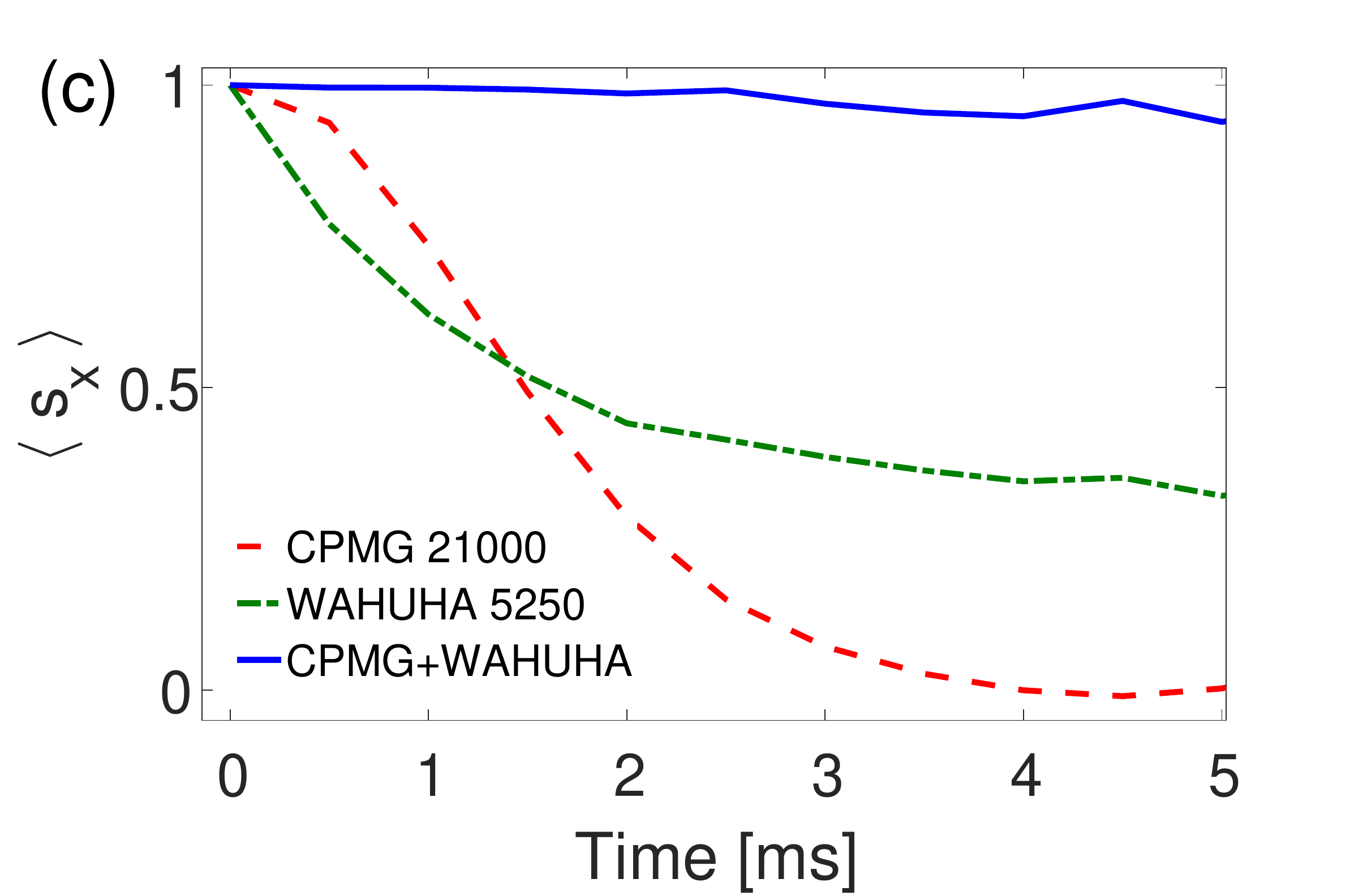}
	\caption{(Color online) Cluster-based simulations of the spin dynamics of a spin ensemble, under DD and the WAHUHA sequence consisting of 1000 pulses. The dominant interactions are: (a) A spin-bath environment, correlation time $\tau_c=5$ $\mu$s and coupling strength $b=20$ kHz, and (b) dipolar interactions among the spins in the ensemble, with a typical interaction strength of 60 Hz (464 spins within a $\approx 4.5$ $\SI{}{\micro\meter}^2$ measurement surface). (c) Combined interaction of spin-bath and dipolar couplings with the same parameter is in (a)-(b), utilizing a total number of $\sim 21000$ pulses: total $21000$ CPMG, WAHUHA, and a combination of 5 WAHUHA within $1000$ CPMG pulses, averaging only 5 dipolar realizations.}
	\label{fig:pulsed}
\end{figure}

We now consider the effects of pulsed DD sequences on the spin dynamics dominated by these interaction sources. Using such sequences with optimally chosen phases along the Bloch Sphere, arbitrary spin states of the ensemble could be preserved  \cite{Farfurnik2015}, and their modification could contribute to the engineering of unique interaction Hamiltonians, potentially leading to the creation of useful non-classical states of the spin ensemble \cite{Cappellaro2009}. We study the effect of the CPMG sequence, consisting of $\pi$-pulses applied along the initialization axis, the similar XY8 DD sequence, which is more robust to pulse imperfections, as well as the WAHUHA sequence, which was designed to decouple dipolar interactions of spin-$1/2$ systems (Fig. \ref{fig:pulsed}). As expected, in the ideal case with no pulse imperfections, the phases of the pulses do not affect the decoupling efficiency, and the CPMG and XY8 sequences produce similar results \cite{Khodjasteh2005,Ryan2010,Wang2012a,Farfurnik2015}. While these DD sequences dramatically improve coherence in a spin-bath dominated environment [Fig. \ref{fig:pulsed}(a)], our simulations show that they do not affect dipolar interactions within the ensemble at all [Fig. \ref{fig:pulsed}(b)]. However, by applying 100 repetitions of the WAHUHA sequence (a total of 400 pulses), dipolar interactions of 60 Hz are decoupled up to a timescale of $50$ ms, which could be observed at cold temperatures. \cite{Jarmola2012,Farfurnik2015}. Similar results for other dipolar interaction strengths (Appendix C) demonstrate that the typical decay time under WAHUHA with 100 repetitions is an order of magnitude longer than the typical dipolar interaction time, and more repetitions of this sequence will result in even higher fidelities. 
\paragraph{}
Since the WAHUHA sequence is not efficient for decoupling the ensemble from the spin-bath environment [Fig. \ref{fig:pulsed}(a)], in order to achieve full decoupling in a realistic scenario with both types of interactions, one has to combine DD pulses and WAHUHA. Figure \ref{fig:pulsed}(c) depicts the dynamics under a combined sequence, for which 5 WAHUHA repetitions are applied between every adjacent pair of CPMG $\pi$-pulses, compared to the application of the same amount (21,000) of CPMG or WAHUHA pulses only. In contrast to the results of Fig. \ref{fig:pulsed}(a)-(b), which were completely dominated by a single interaction source, in the scenario of Fig. \ref{fig:pulsed}(c), disregarding their different final baselines, both WAHUHA and CPMG result in comparable decay timescales, indicating that spin-bath and dipolar interactions have a nearly-equal contribution. These results with a total of 21,000 applied pulses, demonstrate that such a combination could lead to the preservation of the spin state up to $5$ ms, while the application of DD alone ($\pi$-pulses) or WAHUHA alone is no longer effective. The modification of such sequences could lead to the creation of useful non-classical states of the ensemble \cite{Cappellaro2009}. For a given sample, if the origin of the dominant interactions of the ensemble is not known (internal dipolar/spin-bath), the separate application of both sequences could identify this dominant term: a train of $\pi$-pulses significantly enhances the fidelity only for interactions dominated by the bath, while WAHUHA significantly enhances the fidelity only when the internal dipolar interactions dominate.
\subsection{D. Finite Width Pulses}
\begin{figure}[!t]	
	\includegraphics[width=1\columnwidth]{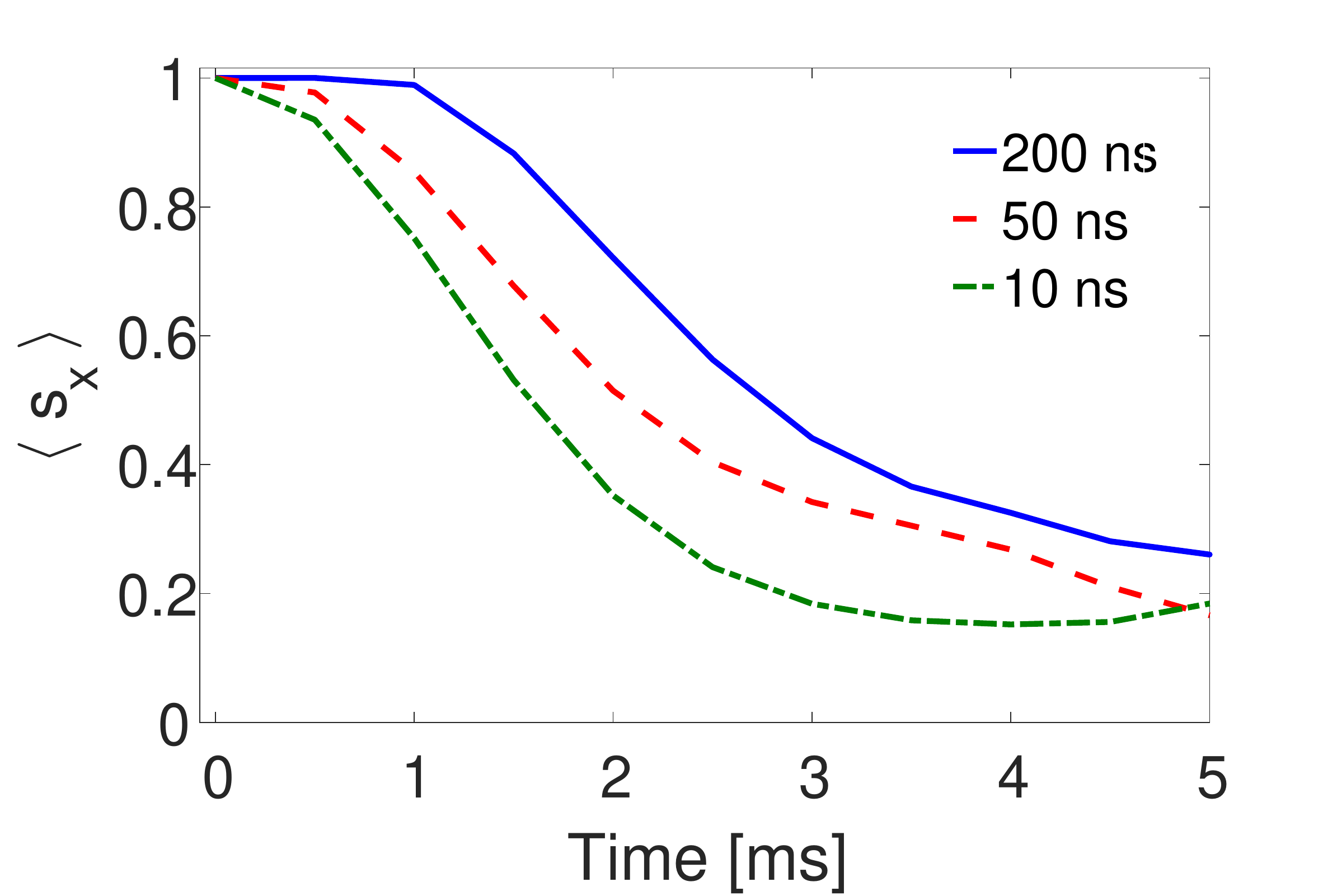} 
	\includegraphics[width=1\columnwidth]{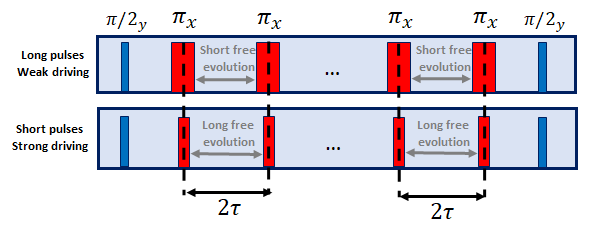} 
	\caption{(Color online) Cluster-based simulations of the spin dynamics of an ensemble consisting of 464 spins, under a realistic spin-bath environment and dipolar coupling (same parameters as before), utilizing 4000 realistic CPMG pulses with various finite durations. Since weaker driving results in shorter free evolution times for dephasing between pulses, the spin state is better-initialized along the driving axis, which leads to a more efficient decoupling of the MW driving from dipolar interactions and longer decay times.}
	\label{fig:duration}
\end{figure}

Finally, we consider the effect of $\pi$-pulses with realistic finite durations on the spin dynamics of the ensemble utilizing the CPMG and XY8 protocols (Fig. \ref{fig:duration}). 
 When considering spin-bath only, we find that the spin dynamics are not affected by varying pulse durations. However, the decay time significantly grows with the pulse durations in realistic scenarios for which spin-bath and internal dipolar interactions exist. We explain this outcome in the following way: sufficiently strong MW driving decouples from the bath completely, but it is efficient for decoupling dipolar interactions only when the state is initialized along the driving axis. While applying a DD sequence, the spin state is no longer initialized along the same axis, both due to coherent dipolar dynamics and dephasing caused by the interactions with the bath, within the free evolution times between pulses. For longer pulses, the free evolution times are shorter, leading to a better preservation of the state initialized along the driving axis before the beginning of the next pulse. This results in an enhanced decoupling over the case of short pulses, in which dephasing within the free evolution times is more significant. This effect is more significant for the CPMG sequence (Fig. \ref{fig:duration}), in which the driving is always along the initialization axis, than for the XY8 sequence (Appendix C), in which the driving is along this axis only half of the time. 
\section{ IV. Conclusions}
To summarize, by simulating the dynamics of an ensemble of 464 spins in a spin-bath environment using a cluster approach and an exact OU algorithm, we showed that a strong enough spin-lock driving (two orders of magnitude stronger than the interactions) could decouple the related interactions for states initialized along the driving axis. The separate application of the CPMG and WAHUHA sequences could identify the dominant interaction source, while their combined application could decouple both types of interactions, preserving the spin state of the ensemble. Additional modification of such sequences could lead to the generation of engineered interaction Hamiltonians, creating non-classical states of the ensemble, and contributing to quantum sensing and quantum information prcessing \cite{Cappellaro2009}. Finally, the duration of the applied pulses may affect the decoupling efficiency, due to imperfect initialization within the free evolution times, as well as the interplay between the spin-bath, internal dipolar interactions and the MW driving within the durations of the pulses.
\section{acknowledgements}
We thank Yonatan Hovav, Nati Aharon, Connor Hart, Erik Bauch, Jennifer M. Schloss, Matthew Turner, Emma Rosenfeld, Ronald L. Walsworth, Joonhee Choi, Hengyun Zhou, Mo Chen and Paola Cappellaro for the fruitful discussions. This work has been supported in part by the Minerva ARCHES award, the CIFAR-Azrieli global scholars program, the Israel Science Foundation (grant No. 750/14), the Ministry of Science and Technology, Israel, and the CAMBR fellowship for Nanoscience and Nanotechnology.

\appendix
\section{Appendix A: Two-level subspace of spin one dipolar interactions}
\paragraph{}
 When MW driving is applied resonant with the $m_s= 0\leftrightarrow+1$ transition (for example), the spin manifold can be treated as a two-level subspace of the spin-triplet \cite{Taylor2008}. The effective Hamiltonian representing an ensemble of such spins interacting with each other by dipolar interactions is given by \eqref{eq:dipolar}. Note that expression \eqref{eq:dipolar} slightly differs from the interaction Hamiltonian among spin-1/2 particles, for which the coefficient of $\sigma^z_i\sigma^z_j$ is $-3$. Although this difference may seem incremental, it can lead to observations of unique physical phenomena when the Hamiltonian is modified by pulsed MW control. For example, the WAHUHA sequence \cite{Waugh1968,Rhim1973,Mansfield1971}, consisting of four $\frac{\pi}{2}$-pulses with unequal timings, was designed to decouple spin-1/2 dipolar interactions. 
\begin{figure*}[!t]	
	\includegraphics[width=2\columnwidth]{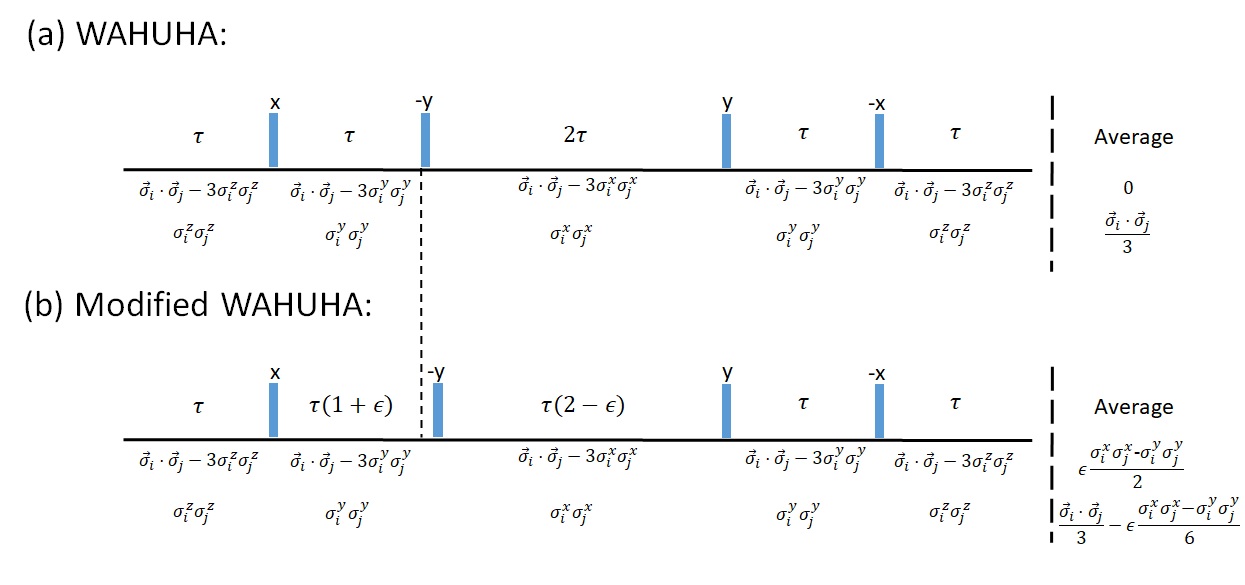} 
	\caption{(Color online) Average Hamiltonian calculation of a spin-1/2 ensemble dipolar Hamiltonian, and the addition term in the case of two level subspace of a spin one system, under conventional and modified WAHUHA sequences.}
	\label{fig:WAHUHA}
\end{figure*}
Indeed, in the first order average Hamiltonian picture \cite{Waugh1968}, the effective Hamiltonian of such interactions yields zero. In the NV-NV dipolar interaction Hamiltonian, however, the extra $\sigma^z_i\sigma^z_j$ term results in the effective Hamiltonian $\frac{\vec{\sigma}_i \cdot \vec{\sigma}_j}{3}$ [Fig. \ref{fig:WAHUHA}(a)].

Such an additional term can contribute to the creation of robust non-classical states of the spin ensemble \cite{Rey2008,Cappellaro2009}. For example, if one of the WAHUHA pulses is slightly shifted by a small time increment $\epsilon \tau$ [Fig. \ref{fig:WAHUHA}(b)],  the resulting interaction Hamiltonian yields $\epsilon \frac{\sigma^x_i\sigma^x_j-\sigma^y_i\sigma^y_j}{2}$ which, under certain conditions, can generate spin-squeezed states $45^0$ along the x-y plane of the Bloch Sphere \cite{Cappellaro2009}. Such states, not yet demonstrated in the solid state, could eventually break the standard quantum limit (SQL), leading to novel directions in quantum sensing. However, in order for these generated non-classical states to remain robust, the total angular momentum number $J^2$ has to be conserved. Such a condition is fulfilled when the squeezing-generating Hamiltonian acts as a small perturbation to another $J^2$-conserving Hamiltonian \cite{Rey2008,Cappellaro2009}. In our case, the effective Hamiltonian  $\frac{\vec{\sigma}_i \cdot \vec{\sigma}_j}{3}$, appearing in the unique case of a two-level subspace of spin one systems such as NV ensembles, acts as such an $J^2$-conserving terms. This highlights the potential of these systems for the generation of robust non-classical states, over well-studied spin-1/2 systems. It is therefore interesting to estimate the dynamics of such systems, which is the subject of this work.

\section{Appendix B: Simulation Methods}
In this Appendix, we present our simulation methods and justify their validity for evaluating the exact dipolar dynamics of systems up to 10 spins, as well as the expected dynamics for large (hundreds or thousands) numbers of spins. Due to computational cost limitations, the dynamics of large spin ensembles cannot be simulated explicitly from the interaction Hamiltonian, our method is cluster-based \cite{Maze2008b}. Our analysis below, exhibiting convergence of the dynamics with the use clusters consisting of six spins, emphasizes the effectiveness of the simulation method in estimating the dynamics of large ensembles. The Appendix is organized as follows: In section (i), we consider a simplified (non-realistic) model of all-to-all equal couplings to obtain an analytical expression for the spin dynamics, which can be solved analytically to provide intuition for the simulated dynamics later on. Section (ii) describes our method of cluster simulations and its obtained results for the dynamics for different cluster sizes in the realistic case of random spin positions. Using Fourier and convergence analysis on driven and non-driven dynamics, the convergence of six-cluster spins is justified in section (iii).
\subsection{(i) Analytical all-to-all equal interactions model}
\begin{figure}[!t]	
	\includegraphics[width=1\columnwidth]{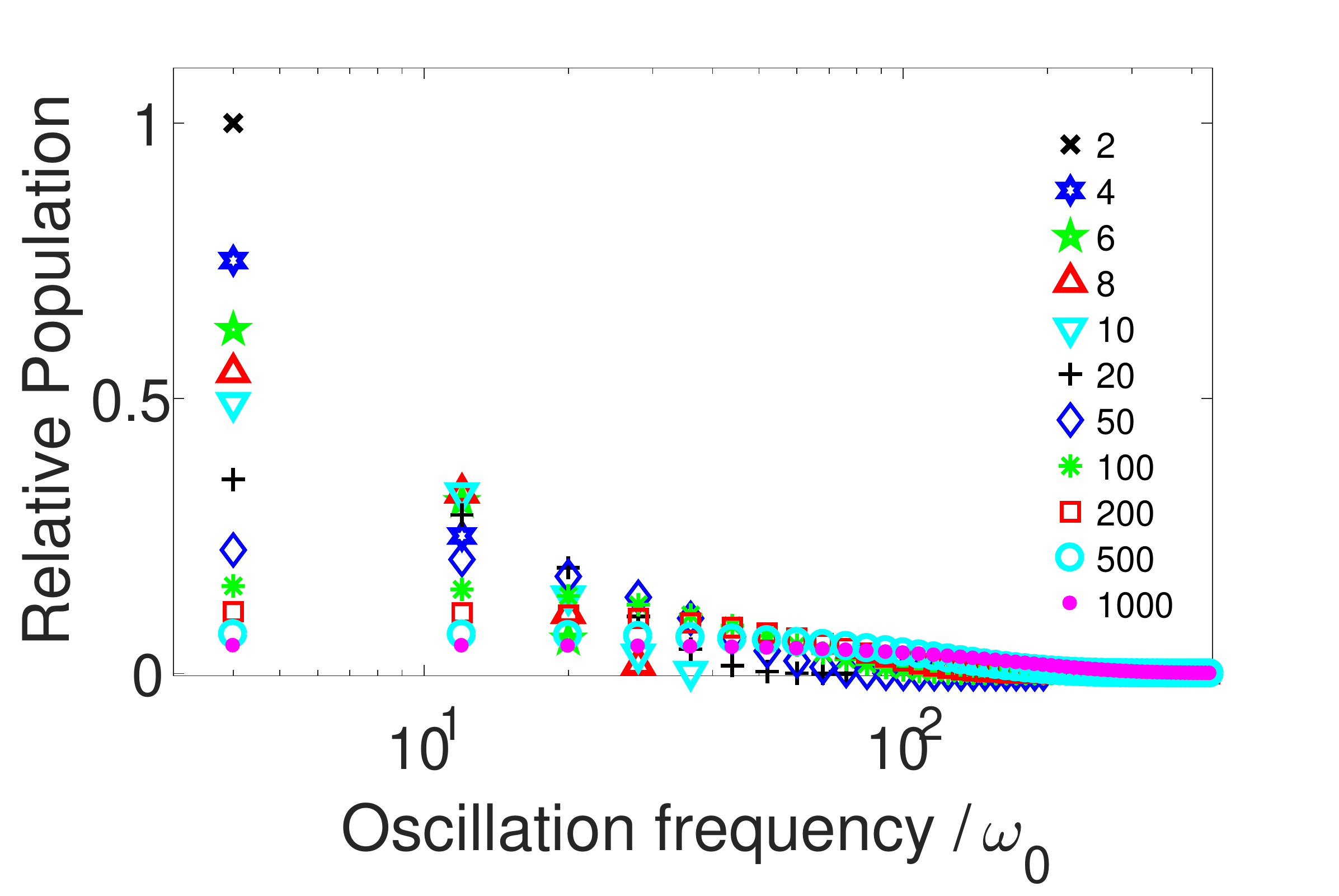} 
	\caption{(Color online) Oscillation frequencies and their populations for different numbers of spins in the limit of couplings with equal strength $\omega_0$, calculated from eq. \eqref{eq:Sx2}. Larger numbers of spins introduce additional higher frequencies, and a significant drop in the populations of all terms.}
	\label{fig:equal}
\end{figure}

\paragraph{}
In the simplest (and non-realistic) scenario, we consider an ensemble consisting of n spins with all-to-all dipolar interactions with equal strengths $\omega_{ij} \equiv \omega_0$. All spins are initialized to the same (``x") axis in the Bloch sphere,
\begin{equation}
|+\rangle=\frac{1}{\sqrt{2}^n} \sum^n_{k=0} |\uparrow^{(k)}\downarrow^{(n-k)}\rangle,  
\label{eq:upx}
\end{equation} where $|\uparrow^{(k)}\downarrow^{(n-k)}\rangle$ denotes a sum over all ${n}\choose{k}$ combinations in which $k$ spins point up and $n-k$ spins point down. Since the left term in eq. \eqref{eq:dipolar} is isotropic, the dynamics is dominated by the Hamiltonian $H_0=-2 w_{0}\sum_{i,j}\sigma^z_i\sigma^z_j$. For a given value of $k$ spins pointing up, there are $k(n-k)$ different combinations in which spins $i$, $j$ have opposite signs such that $\sigma^z_i\sigma^z_j\rightarrow -1$, and ${n\choose2}-k(n-k)$ combinations in which these spins share the same sign, such that $\sigma^z_i\sigma^z_j\rightarrow 1$. Therefore, the application of the time evolution operator $U=e^{-iH_0 t}$ on the initial state yields 
\begin{equation}
|\psi(t)\rangle=\frac{1}{\sqrt{2}^n} \sum^n_{k=0} e^{2i\omega_0 t \left[{{n}\choose{2}}-2k(n-k)\right]}|\uparrow^{(k)}\downarrow^{(n-k)}\rangle.  
\label{eq:wavefunc}
\end{equation}
Let us now calculate the spin polarization along the initialization axis, $\langle S_x \rangle = \frac{1}{n}\sum_k \langle S_{xk} \rangle = \frac{1}{n}\left(\langle \sigma_x\otimes \mathbb{I}\otimes \cdots \otimes\mathbb{I}\rangle+\cdots +\langle\mathbb{I}\otimes \mathbb{I}\otimes \cdots \otimes \sigma_x\rangle\right)$. Since all spins are equivalent, it is sufficient to calculate the expectation value of one spin operator only, $\langle S_x \rangle=\langle S_{x1} \rangle=\langle \left(|\uparrow\rangle \langle \downarrow|+|\downarrow \rangle \langle \uparrow|  \right)\otimes \mathbb{I}\otimes \cdots \otimes\mathbb{I}\rangle$. After applying $|\psi(t)\rangle$ from both sides, this expectation value yields 
\begin{equation}
\langle S_x\rangle=\frac{1}{2^{n-1}} \left(e^{4i\omega_0 t (n-1)}+\sum^{n-1}_{k=1} {{n-1}\choose{k-1}} e^{4i\omega_0 t (2k-n-1)}\right),  
\label{eq:Sx1}
\end{equation}
where the first term represents the contribution from the states in which all spins point up / down, the sum represents the states in which $k$ spins point up (given that the first spin points up, ${n-1}\choose{k-1}$ is the number of combinations for which $k-1$ of the other spins point up), and a factor of 2 arises from the symmetry between up and down. Expression \eqref{eq:Sx1} is always real, and can be rewritten for even numbers of spins in the form: 
\begin{equation}
\begin{split}
\langle S_x\rangle=\frac{1}{2^{n-2}} \Bigg[ & \cos\left(4\omega_0 t (n-1)\right)\\ &+\sum^{\ceil{(n-1)/2}}_{k=2} {{n-1}\choose{k-1}} cos\left(4\omega_0 t (2k-n-1)\right)\Bigg],  
\label{eq:Sx2}
\end{split}
\end{equation}
with an additional DC term for odd numbers of spins, $\frac{1}{2^{n-2}}{{n-1}\choose{(n-1)/2}}$. Using the analytical expressions \eqref{eq:Sx1},\eqref{eq:Sx2} the spin dynamics can be easily calculated in the special case of equal couplings up to thousands of spins, in contrast to our ability to perform exact simulations on the Hamiltonian \eqref{eq:dipolar}, which is limited to 10 spins. The resulting oscillation frequencies and their populations, identical to those obtained by a direct simulation with equal couplings, are shown in Fig. \ref{fig:equal}.
For growing numbers of spins, oscillating terms with higher frequencies are added. However, due to the factor $1/2^{n-2}$, the relative population of each oscillating component drops significantly with the number of spins. As will be demonstrated in the following section, this picture does not represent the realistic scenario, in which interactions vary due to the different positions of the spins, thus a cluster approach is necessary.  

\subsection{(ii) Cluster simulations using the explicit Hamiltonian}
\begin{figure}[!t]	
	\includegraphics[width=1\columnwidth]{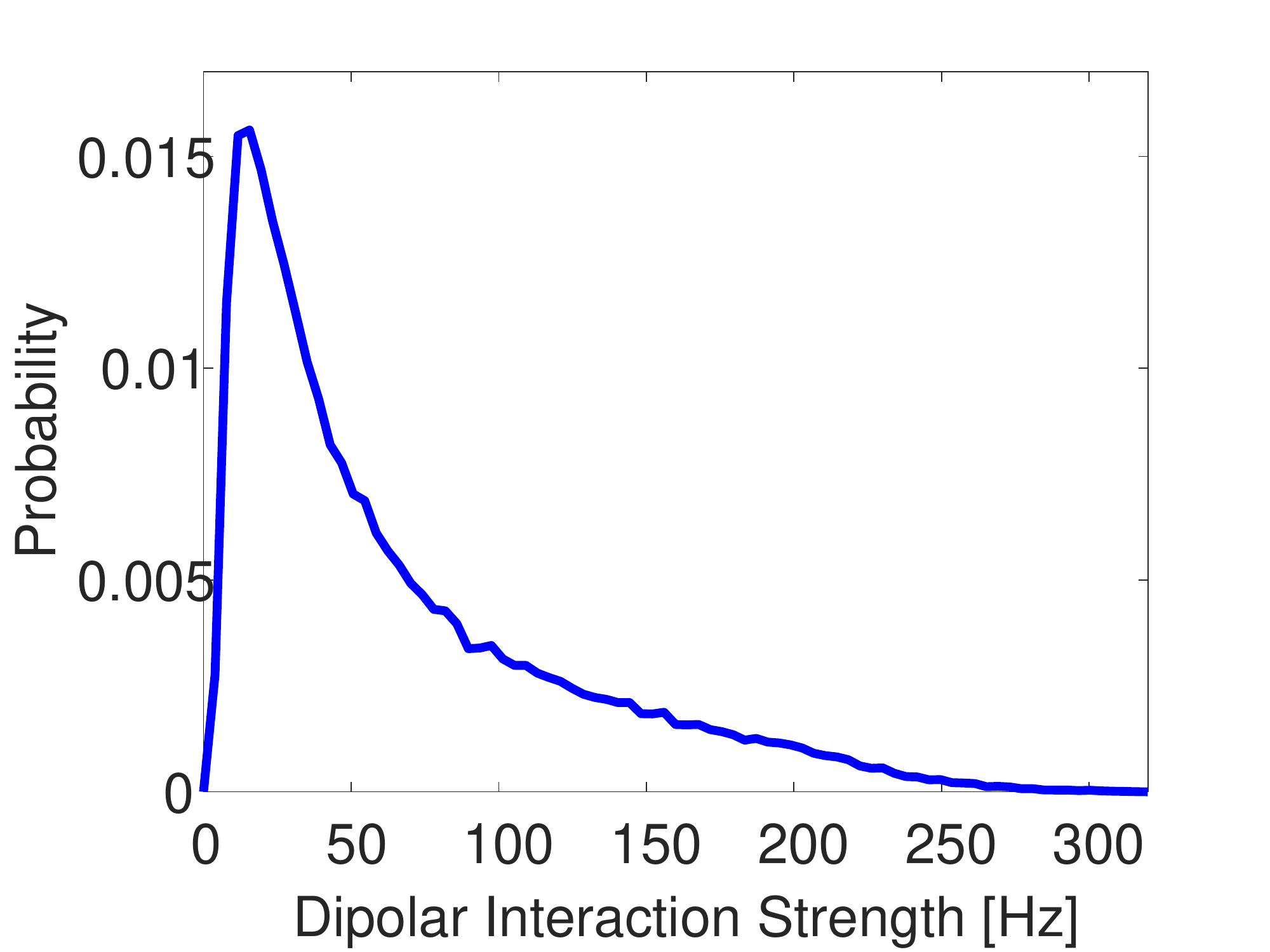} 
	\caption{(Color online) Probability distribution generated for the dominant interactions in a quasi-2D NV ensemble with a spin concentration of $10^{10}$ $\SI{}{\centi\meter}^{-2}$ within a $\approx 4.5$ $\SI{}{\micro\meter}^2$ measurement surface, obtained by averaging over many realizations of randomly-generated 464 spins. The resulting average interaction strength is $\omega_0\approx 60$ Hz.}
	\label{fig:dist}
\end{figure}

\begin{figure}[!t]	
	\includegraphics[width=1\columnwidth]{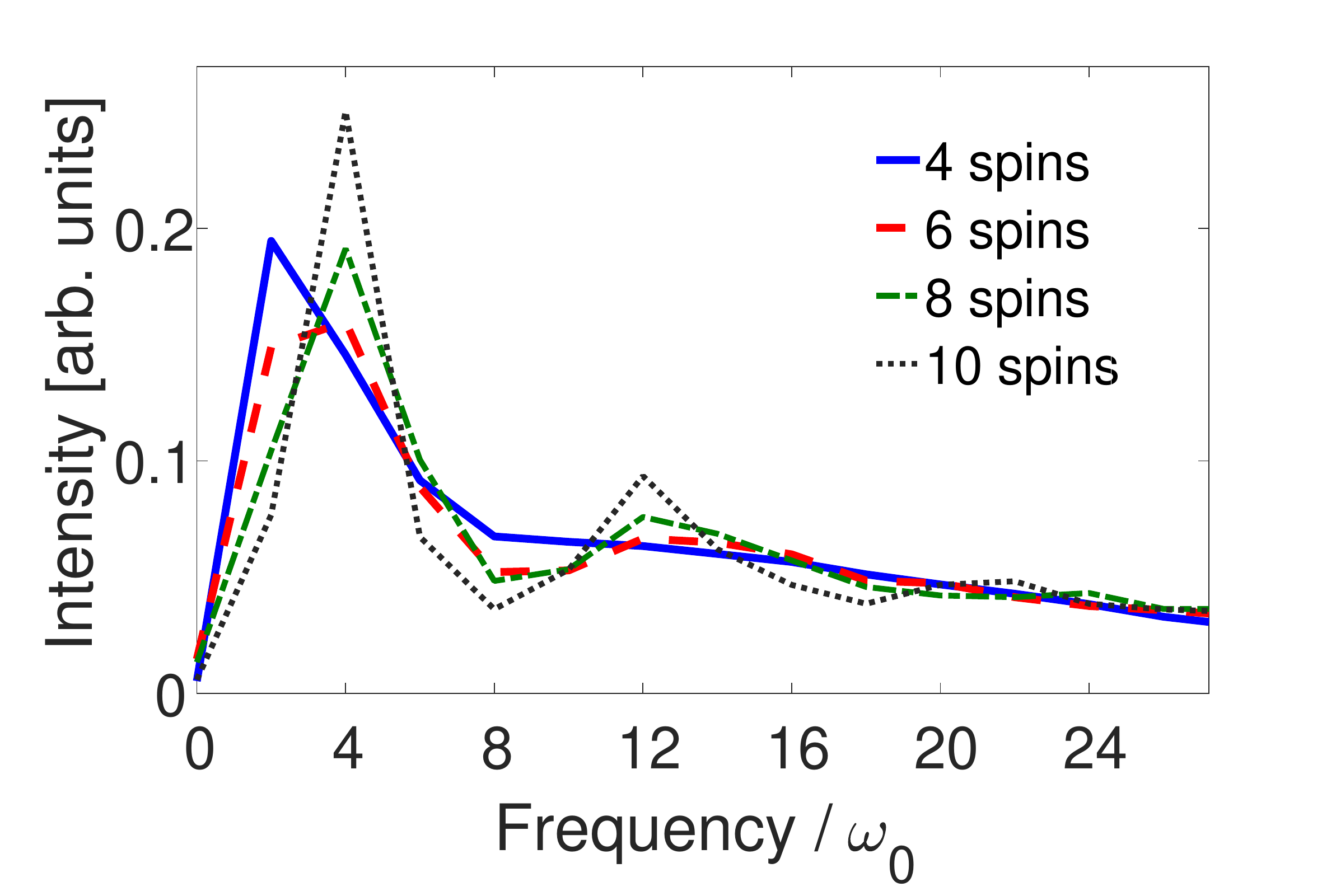} 
	\caption{(Color online) Fourier transform of the dynamics of a spin ensemble consisting of 464 spins with a typical interaction strength $\omega_0\approx 60$ Hz, under the dipolar Hamiltonian \eqref{eq:dipolar}. The interactions within clusters consisting of different numbers of spins are generated from the probability distribution in Fig. \ref{fig:dist}.}
	\label{fig:dipdiffn}
\end{figure}

\paragraph{}
In the previous section we considered a scenario in which all dipolar interactions have equal strengths. This scenario is non-realistic since varying distances between spins correspond to different strengths $\omega_{ij}$. As a result, in order to get a more relevant estimate of the dynamics arising from Hamiltonian \eqref{eq:dipolar}, interaction strengths should be generated from a distribution, which correctly represents the experimental measurement scenario. In our following more-realistic simulations, for an ensemble with a given spin concentration, the proper amount of spins is generated inside the measurement surface / volume, and the interaction of each spin with its nearest neigbor is taken into account. This process is repeated for many realizations until convergence, to form a histogram representing the interaction distribution. Such a typical distribution, considering a quasi-2D NV ensemble with a spin concentration of $10^{10}$ $\SI{}{\centi\meter}^{-2}$, which can be achieved by TEM irradiation \cite{Farfurnik2017}, within a $\approx 4.5$ $\SI{}{\micro\meter}^2$ measurement surface (total 464 spins) is shown in Fig. \ref{fig:dist}. 
\paragraph{}
In order to gain some intuition for the simulated results, Fourier transforms of the dipolar dynamics considering different cluser sizes [Fig. \ref{fig:dipolar}(a)] are shown in  Fig. (\ref{fig:dipdiffn}).
The oscillating frequencies presented in these dynamics are compatible with those obtained from the equal-coupling model [eq. \eqref{eq:Sx2} and Fig. \ref{fig:equal}], with the two most significant terms centered at $4\omega_0$ and $12\omega_0$. Note that the uncertainty in the main frequency $4\omega_0$ is $\pm \omega_0$, which enables to extract the typical interaction strength up to an accuracy of $25\%$. However, in contrast with the equal coupling calculation, the varying spin positions randomly generated from the measurement surface result in the broadening of the frequency peaks, as well as in the significant drop in the contribution of high-frequency terms. For example, the population ratio between the two slowest frequencies is $\sim 2.5$ for the explicit simulations and only $\sim 1.5$ for the equal-coupling model considering 10 spins, emphasizing that the contribution of high frequency terms is greatly diminished in the realistic case. Therefore, we conclude that in spite of the correct prediction of the oscillation frequencies, the equal-coupling model is not suitable for simulating the realistic dynamics for large spin ensembles, which indeed requires the use of a cluster approach. 
	\subsection{(iii) Determining the cluster size}
\begin{figure}[!t]	
		\includegraphics[width=1\columnwidth]{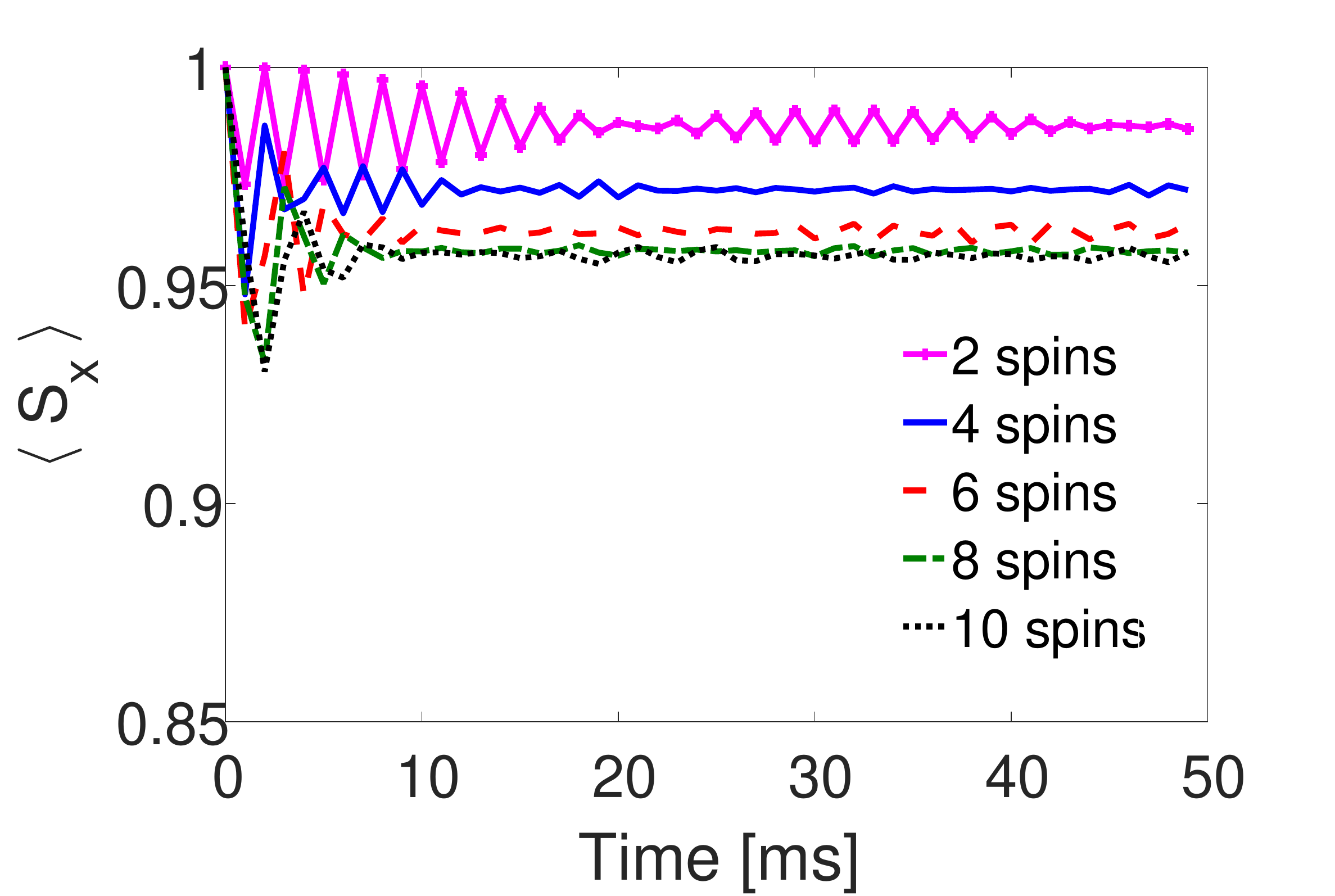} 
		\caption{(Color online) Cluster-approximation simulations of the dynamics of a spin ensemble consisting of 464 spins with a typical interaction strength $\omega_0\approx 60$ Hz , under a spin-lock driving with an intensity of $\Omega\approx 12\omega_0$. The interactions within clusters consisting of different numbers of spins are generated from the probability distribution in Fig. \ref{fig:dist}.}
		\label{fig:SL}	
	\end{figure}

	\paragraph{}
	The sufficient amount of spins in a cluster for estimating spin ensemble dynamics can be estimated by considering the population ratio between different frequency terms, and justified by convergence analysis on driven spin dynamics. The ratios between the three main frequencies in the explicit dynamics (Fig. \ref{fig:dipdiffn}) are $\sim 1:2.5:6$ which, even in the simplified model of equal couplings, is mostly compatible with the choice of six-spin clusters, exhibiting the ratios $\sim 1:2:10$. Although in general, the equal coupling model is not suitable for describing the realistic scenario, our choice of six-spin clusters exhibits the same main oscillation frequencies dominating the dynamics. However, such a comparison is rather rough, and the choice of cluster size has to be justified quantitatively. Indeed, we further justify this chosen number of spins in a cluster by performing convergence analysis on the dynamics of a driven system: the spin dynamics under Hamiltonian \eqref{eq:dipolar} is simulated together with a continuous driving along the initialization axis (spin-lock), which is an order of magnitude stronger than the typical interaction strength, $\Omega\approx 12 \omega_0$ [Fig. \ref{fig:SL}].
	Theoretically, such a driving is expected to cancel out effects of components oscillating slower than the driving strength \cite{Hirose2012}. In the case of Fig. \ref{fig:dist}, a polarization of $\approx 0.95$ is achieved by driving the system with $\Omega\approx 12\omega_0$, which is expected to result in the cancelation of the two first frequency terms $4\omega_0$ and $12\omega_0$. These results charactarize the natural physical dynamics under the dipolar interaction, which will remain the same for any external control applied, making our choice of spins within clusters relevant for simulating any sequence.
	The difference between this converged value and the steady-state polarization obtained by considering clusters of six spins is only about 1 percent, indicating that such a cluster size is suitable for simulating large ensembles. Furthermore, for such a cluster size, even the equal-coupling model (Fig. \ref{fig:equal}) provides a close prediction for the degree of polarization: the population of the first two frequencies sums up to $0.9375$, very close to the converged degree of polarization, indicating that such a choice for the cluster size incorporates the effects of randomness caused by varying dipolar interaction strengths. 
	
\section{Appendix C: Supplemental simulation results}
\subsection{(i) Dipolar interactions for different spin concentrations}
\paragraph{}
The simulations in this work were performed for several different experimental scenarios (concentrations and number of spins): spins of an ensemble with a defined concentration were generated randomly within the a defined measurement surface. As can be seen in Fig. \ref{fig:dipolar}(b), as well as Fig. \ref{fig:diffdist}(b) and Fig. \ref{fig:WAHUHAdiff} below, the qualitative dynamics and decoupling results for diffrerent spin concentrations are similar, with the timescales changing according to the coupling strength ratios. Small Quantitative differences [Fig. \ref{fig:diffdist}(b)] correspond to slightly different structures from the interaction distributions [Fig. \ref{fig:diffdist}(a)].  
\begin{figure}[!t]	
	\includegraphics[width=1\columnwidth]{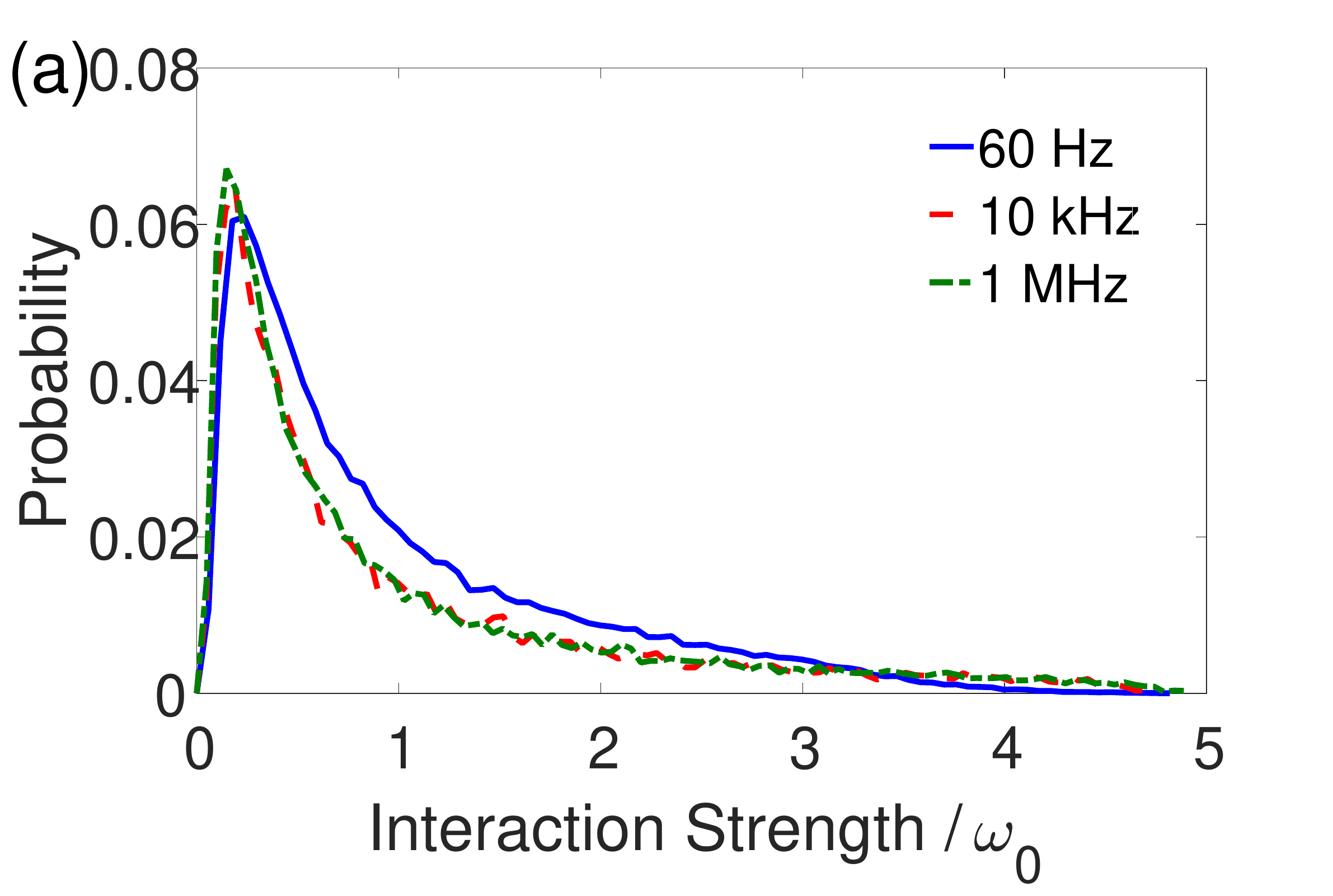} 
	\includegraphics[width=1\columnwidth]{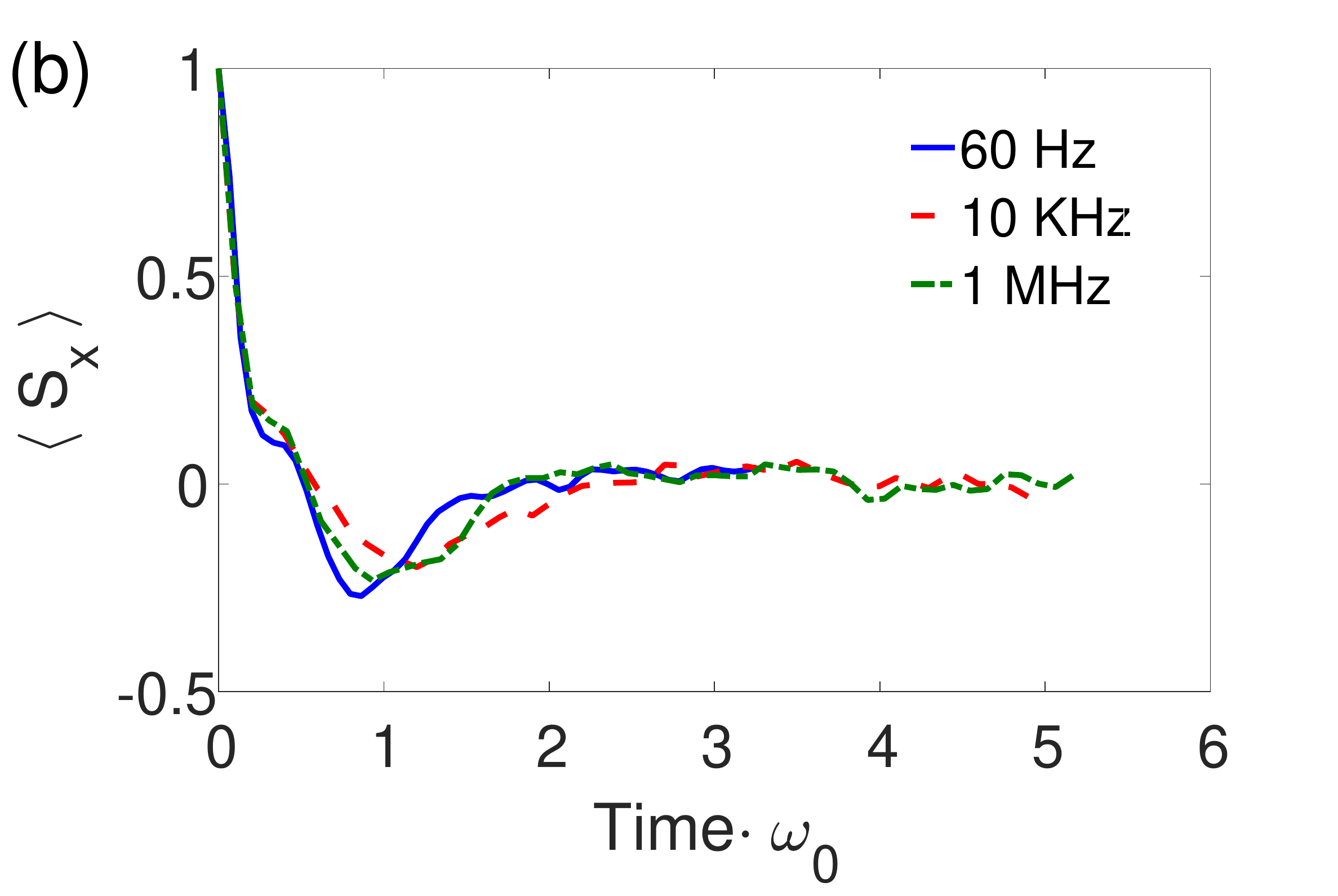} 
	\caption{(Color online) (a) Interaction strength distributions and (b) cluster-based simulations of the spin dynamics of spin ensembles with different properties: Spin concentration of $10^{10}$ $\SI{}{\centi\meter}^{-2}$ within a $\approx 4.5$ $\SI{}{\micro\meter}^2$ measurement surface, representing 464 spins with a typical interaction strength of $\omega_0 \approx 60$ Hz, $\omega_0\approx 10$ kHz typical interaction  within a $\approx 4.5$ $\SI{}{\micro\meter}^2$ measurement surface (9980 spins), and $\omega_0 \approx 1$ MHz within a $\approx 0.46$ $\SI{}{\micro\meter}^2$ measurement surface (9980 spins). The horizontal axes are normalized to $\omega_0$ to emphasize similarities between the results, with small differences due to the slightly-different strength distributions.}
	\label{fig:diffdist}
\end{figure}
\subsection{(ii) Spin-locking on dipolar interaction of $60$ Hz}
\begin{figure}[!t]	
	\includegraphics[width=1\columnwidth]{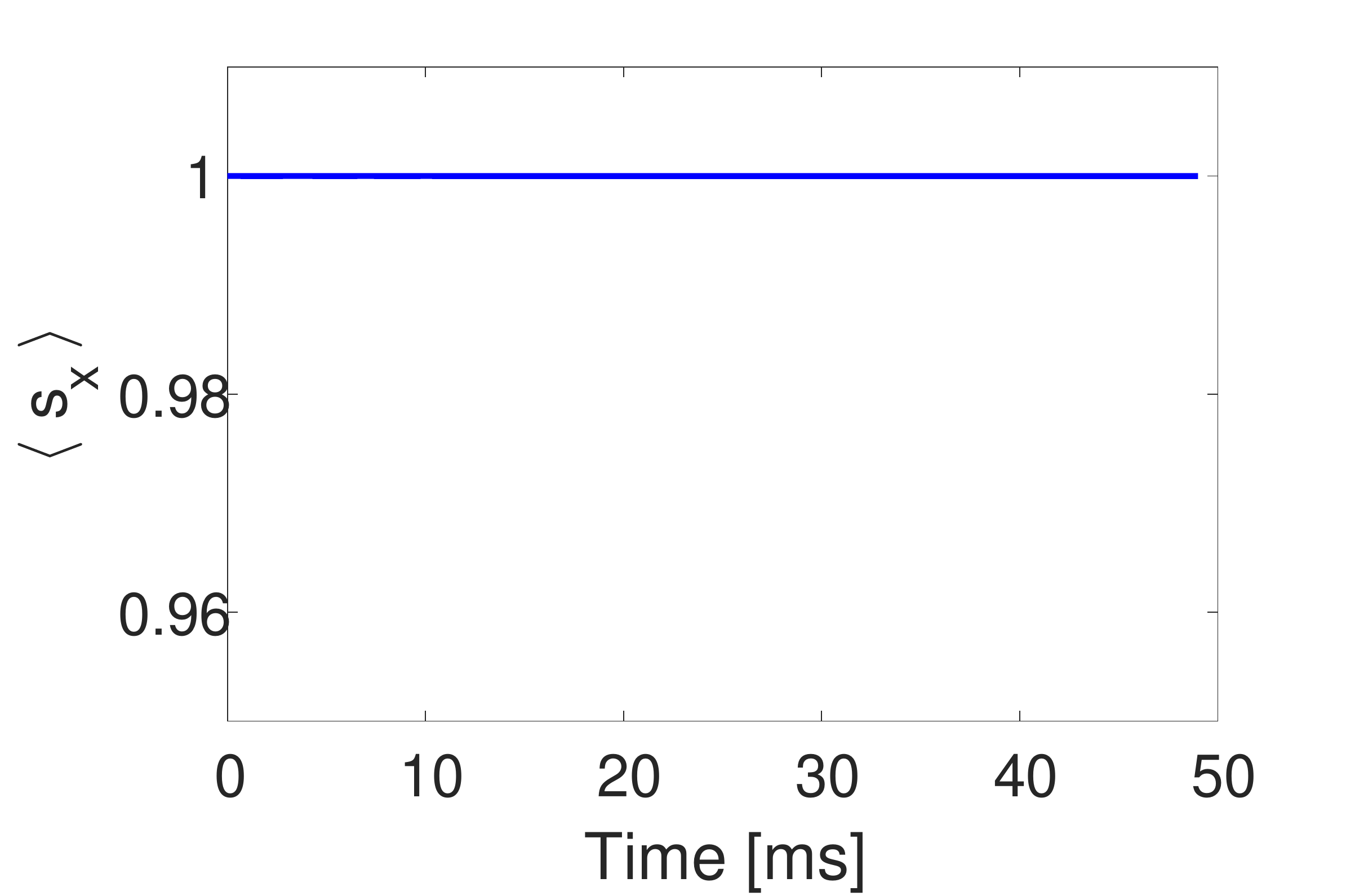}
	\caption{(Color online) Cluster-based simulations of the spin dynamics of an ensemble consisting of 464 spins, under spin-lock driving at with the strength of 0.1 MHz, for dipolar NV-NV interactions with a typical interaction strength of 60 Hz.}
	\label{fig:SL60}
\end{figure}

\paragraph{}
Figure \ref{fig:SL60} demonstrates the dynamics of an NV ensemble with a typical dipolar NV-NV interaction strength of 60 Hz under spin-lock driving with the strength of 0.1 MHz. In consistence with the results given in the main text for other intensities, driving two orders of magnitude stronger than the typical dipolar interactions decouples the interactions completely and results in unity evolution for over more than 10 ms.
\subsection{(iii) WAHUHA on ensembles with different dipolar coupling strengths}
\begin{figure}[!t]	
	\includegraphics[width=1\columnwidth]{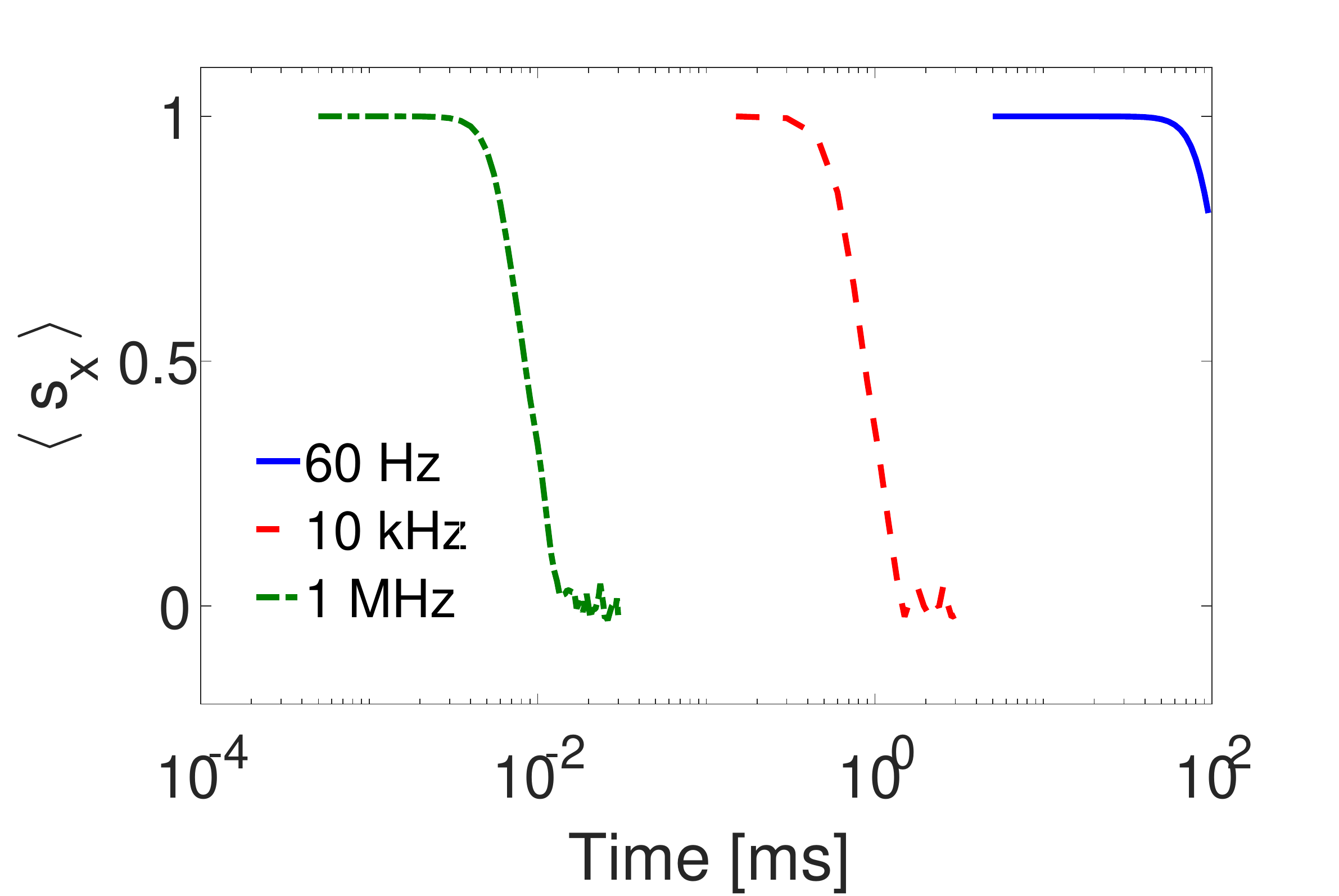}
	\caption{Cluster-based simulations of the spin dynamics under the application of the WAHUHA sequence with 100 repetitions on a spin ensemble dominated by NV-NV dipolar interactions at various concentrations: Spin concentration of $10^{10}$ $\SI{}{\centi\meter}^{-2}$ within a $\approx 4.5$ $\SI{}{\micro\meter}^2$ measurement surface, representing 464 spins with a typical interaction strength of $\omega_0 \approx 60$ Hz, $\omega_0\approx 10$ kHz typical interaction  within a $\approx 4.5$ $\SI{}{\micro\meter}^2$ measurement surface (9980 spins), and $\omega_0 \approx 1$ MHz within a $\approx 0.46$ $\SI{}{\micro\meter}^2$ measurement surface (9980 spins).}
	\label{fig:WAHUHAdiff}
\end{figure}

\paragraph{}
Figure \ref{fig:WAHUHAdiff} demonstrates the dynamics of NV ensembles with various dipolar NV-NV interaction strengths under the application of 100 repetitions of the WAHUHA sequence. The resulting decay times are one order of magnitude longer than the typical interaction time. Furthermore, a comparison to the spin dynamics without external control [Fig. \ref{fig:dipolar}(b)] shows that this amount of repetitions enhances the decay time by two orders of magnitude over the case without any control. Qualitatively, By utilizing a constant amount of repetitions, the decay structure exhibited in different samples is similar, with the timescales changing according to the coupling strength ratios, while small quantitative differences correspond to slightly-different structures of the interaction distributions.  
\subsection{(iv) XY8 with finite pulse durations}
\begin{figure}[!t]	
	\includegraphics[width=1\columnwidth]{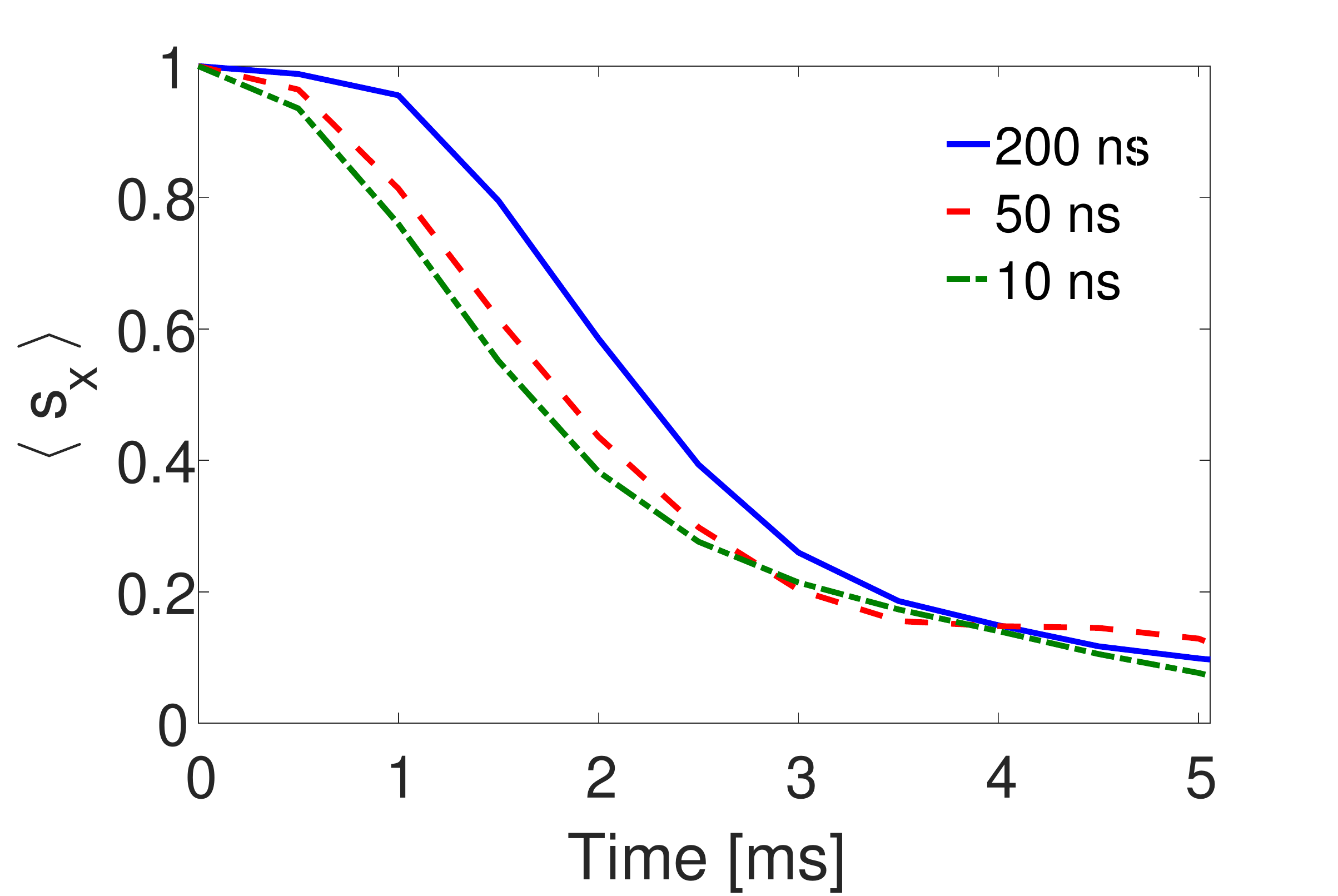} 
	\caption{(Color online) Cluster-based simulations of the spin dynamics of an ensemble consisting of 464 spins, under a realistic spin-bath environment and dipolar coupling (same parameters in Fig. \ref{fig:duration}), utilizing 500 repetitions of the XY8 sequence (total of 4000 pulses) with various finite durations. Since weaker driving results in shorter free evolution times for dephasing between pulses, the spin state is better-initialized along the driving axis, which leads to a more efficient decoupling of the MW driving from dipolar interactions and longer decay times.}
	\label{fig:XY8duration}
\end{figure}

\paragraph{} 
As described in section III D and emphasized in Fig. \ref{fig:duration}, the decoupling efficiency from simultaneous spin-bath and internal dipolar interactions within an NV ensemble might increase with as a function of the pulse durations of the applied DD sequence. Here we show (Fig. \ref{fig:XY8duration}) that this effect is also expressed by utlizing the XY8 sequence. The effect is much more significant for CPMG (Fig. \ref{fig:duration}), where all pulses are applied along the initialization axis, than for XY8 (Fig. \ref{fig:XY8duration}), where only half of the pulses are applied along the initialization axis.  

\subsection{(v) Variations between different realizations in combined spin-bath and dipolar simulations}
\begin{figure}[!t]	
	\includegraphics[width=1\columnwidth]{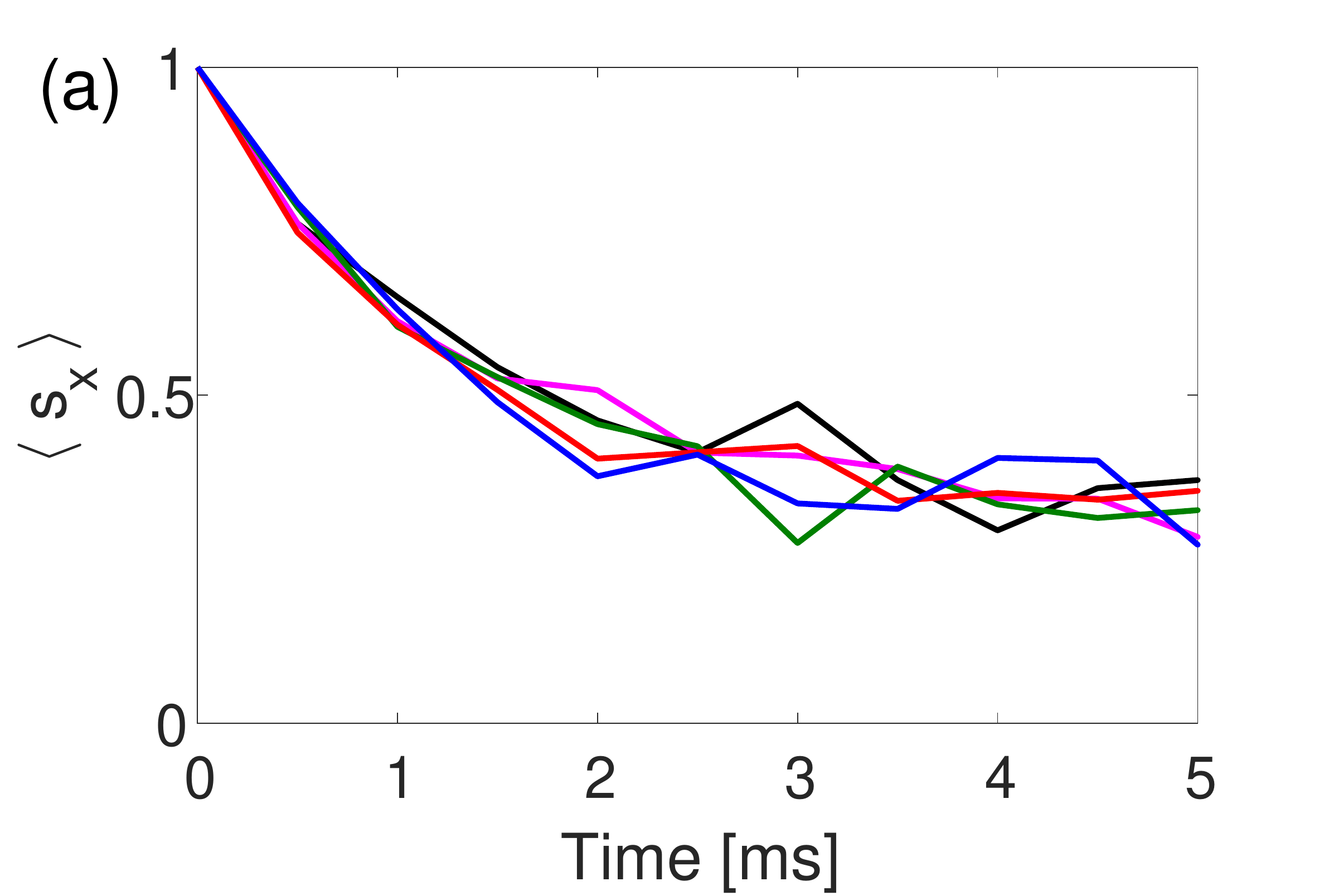} 
	\includegraphics[width=1\columnwidth]{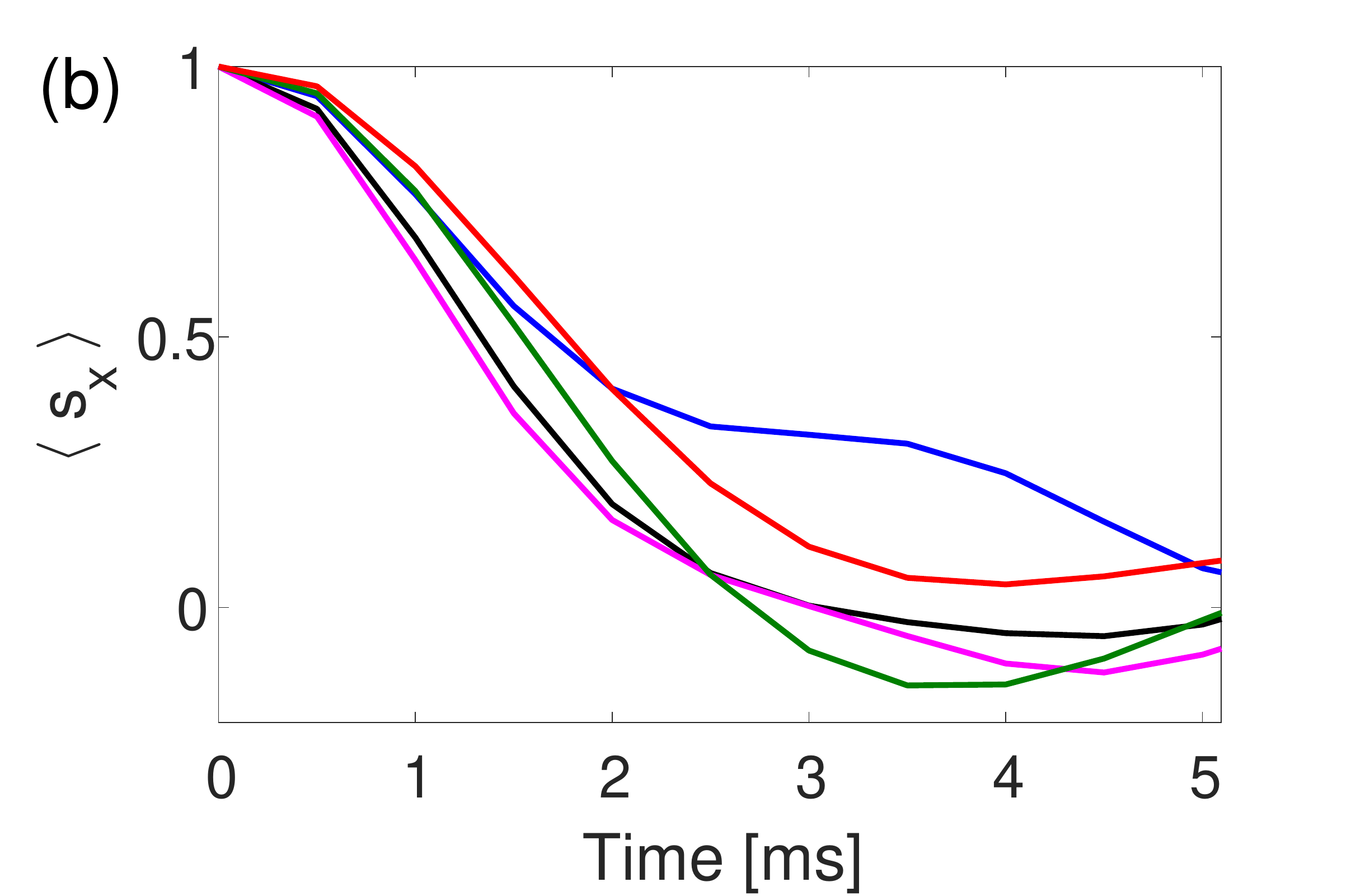} 
	\includegraphics[width=1\columnwidth]{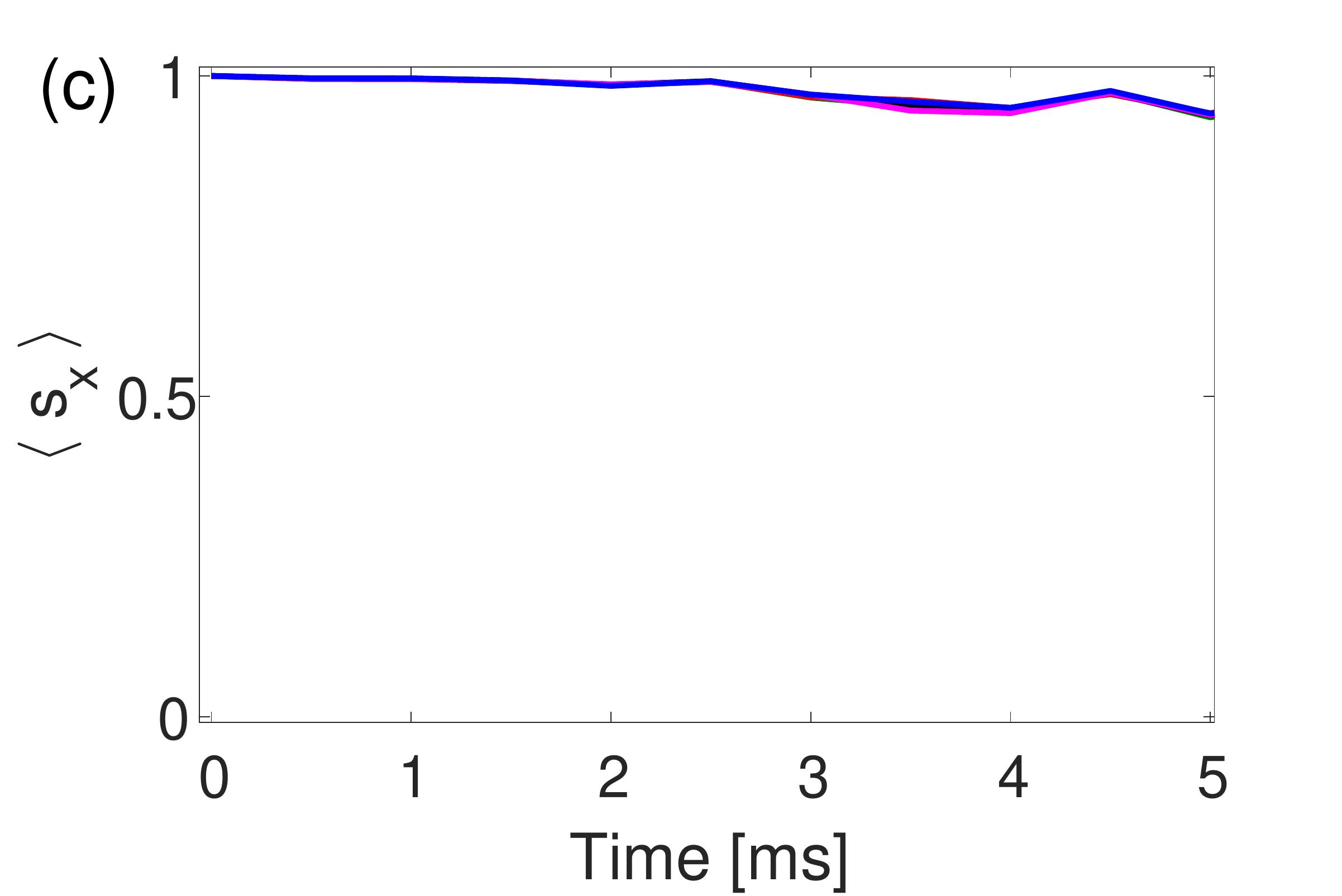} 
	\caption{(Color online) Different realizations for the (a) WAHUHA, (b) CPMG and (C) combined WAHUHA and CPMG expressed in Fig. \ref{fig:pulsed}(c).}
	\label{fig:realizations}
\end{figure}

\paragraph{} 
Since the simulation of OU process requires evolution in small time increments $\Delta t$ and averages over noise realizations, it is very time-consuming, thus for the combined simulations under spin-bath and dipolar interactions [Fig. \ref{fig:spinlock}(c) and Fig. \ref{fig:pulsed}(c)] only five realizations of the dipolar interactions were considered. The explicit results for individual realizations leading to the results of Fig. 3 \ref{fig:pulsed}(c) are presented here in Fig. \ref{fig:realizations}. For the application of WAHUHA [Fig. \ref{fig:realizations} (a)], internal dipolar interactions are fully decoupled up to the simulated timescales, thus different realizations result in the same dynamics dominated solely by interactions with the bath. Since the CPMG sequence does not decouple the effects of internal dipolar interactions, different realizations of different randomly-generated spins in the cluster correspond to different decay scales [Fig. \ref{fig:realizations} (b)]. Qualitatively, however, the total decay trend expressed here and in the main text is very clear even after considering only several realizations. More importantly, by applying a control scheme that simultaneously decouples interactions with the bath and internal dipolar interactions (such as by the combined application of WAHUHA and CPMG expressed in Fig. \ref{fig:realizations} (c)], the resulting dynamics is similar for different realizations.   

\bibliography{nvbibliography}

\end{document}